\documentclass[twocolumn]{aastex701}

\begin{document}

% \title{Template \aastex v7.0.1 Article with Examples\footnote{Footnotes can be added to titles}}
\title{{Low-mass and High-sSFR H$\alpha$ Emitters at $z \sim 2$ in the ZFOURGE--COSMOS Field}}

\author[orcid=0009-0005-5412-4521, sname='Yun']{Jeung Yun}
\affiliation{Department of Astronomy, Graduate School of Science, 
The University of Tokyo, 7-3-1 Hongo, Bunkyo-ku, Tokyo 113-0033, Japan}
\affiliation{Institute of Astronomy, Graduate School of Science, The University of Tokyo, 2-21-1 Osawa, Mitaka, Tokyo 181-0015, Japan}
\email[show]{yun-jeung@g.ecc.u-tokyo.ac.jp}

\author[orcid=0000-0002-0724-9146]{Kentaro Motohara} 
\affiliation{Department of Astronomy, Graduate School of Science, 
The University of Tokyo, 7-3-1 Hongo, Bunkyo-ku, Tokyo 113-0033, Japan}
\affiliation{Institute of Astronomy, Graduate School of Science, The University of Tokyo, 2-21-1 Osawa, Mitaka, Tokyo 181-0015, Japan}
\affiliation{National Astronomical Observatory of Japan, 2-21-1 Osawa, Mitaka, Tokyo 181-8588, Japan}
\email{kmotohara@ioa.s.u-tokyo.ac.jp}

\author[orcid=0009-0000-5163-2305]{Kosuke Kushibiki} 
\affiliation{National Astronomical Observatory of Japan, 2-21-1 Osawa, Mitaka, Tokyo 181-8588, Japan}
\affiliation{Graduate Institute for Advanced Studies, SOKENDAI, 2-21-1 Osawa, Mitaka, Tokyo 181-8588, Japan}
\email{kosuke.kushibiki@nao.ac.jp}

\author[orcid=0000-0002-0486-5242]{Nuo Chen} 
\affiliation{Astronomical Institute, Tohoku University, 6-3, Aramaki, Aoba, Sendai, Miyagi 980-8578, Japan}
\email{nuo.chen@astr.tohoku.ac.jp}

\author[orcid=0000-0003-4907-1734]{Masahiro Konishi} 
\affiliation{Department of Astronomy, Graduate School of Science, 
The University of Tokyo, 7-3-1 Hongo, Bunkyo-ku, Tokyo 113-0033, Japan}
\affiliation{Institute of Astronomy, Graduate School of Science, The University of Tokyo, 2-21-1 Osawa, Mitaka, Tokyo 181-0015, Japan}
\email{konishi@ioa.s.u-tokyo.ac.jp}

\author[orcid=0009-0008-9784-7991]{Hidenori Takahashi} 
\affiliation{Department of Astronomy, Graduate School of Science, 
The University of Tokyo, 7-3-1 Hongo, Bunkyo-ku, Tokyo 113-0033, Japan}
\affiliation{Institute of Astronomy, Graduate School of Science, The University of Tokyo, 2-21-1 Osawa, Mitaka, Tokyo 181-0015, Japan}
\email{nori@ioa.s.u-tokyo.ac.jp}

\author{Tomoya Yukino} 
\affiliation{Department of Astronomy, Graduate School of Science, 
The University of Tokyo, 7-3-1 Hongo, Bunkyo-ku, Tokyo 113-0033, Japan}
\affiliation{Institute of Astronomy, Graduate School of Science, The University of Tokyo, 2-21-1 Osawa, Mitaka, Tokyo 181-0015, Japan}
\email{yukino-tomoya867@g.ecc.u-tokyo.ac.jp}

\author[orcid=0000-0002-0100-1238]{Shuhei Koyama} 
\affiliation{National Astronomical Observatory of Japan, 2-21-1 Osawa, Mitaka, Tokyo 181-8588, Japan}
\email{shuhei.koyama@nao.ac.jp}

\author[orcid=0000-0002-0498-5041]{Akiyoshi Tsujita} 
\affiliation{Department of Astronomy, Graduate School of Science, 
The University of Tokyo, 7-3-1 Hongo, Bunkyo-ku, Tokyo 113-0033, Japan}
\affiliation{Institute of Astronomy, Graduate School of Science, The University of Tokyo, 2-21-1 Osawa, Mitaka, Tokyo 181-0015, Japan}
\email{tsujita@ioa.s.u-tokyo.ac.jp}

\author[orcid=0000-0002-4937-4738]{Ichi Tanaka} 
\affiliation{Subaru Telescope, National Astronomical Observatory of Japan, 650 North A’ohoku Place, Hilo, HI 96720, USA}
\email{ichi@naoj.org}

\author[orcid=0000-0002-6656-1631]{Hirofumi Okita} 
\affiliation{Subaru Telescope, National Astronomical Observatory of Japan, 650 North A’ohoku Place, Hilo, HI 96720, USA}
\email{okita14@naoj.org}

\author[0000-0003-4143-4246]{Tsuyoshi Terai} 
\affiliation{Subaru Telescope, National Astronomical Observatory of Japan, 650 North A’ohoku Place, Hilo, HI 96720, USA}
\email{terai@naoj.org}

\begin{abstract}

We present a near-infrared spectroscopic follow-up study of seven H$\alpha$ emitter candidates at $z\sim2$ in the ZFOURGE--COSMOS field, selected via the $K_{\rm s}$-band flux excess method. Spectra obtained with Subaru/SWIMS confirm strong H$\alpha$ emission line at $z=2.06$--2.49 for all the targets. Three galaxies are low-mass systems ($\log(M_\star/M_\odot) < 9$) with high specific SFRs, while four are high-mass systems ($\log(M_\star/M_\odot) > 10$) on the star-forming main sequence. The low-mass galaxies show consistent H$\alpha$ flux measurements between the $K_{\rm s}$-band flux excess method and the spectroscopy, whereas the high-mass galaxies show higher H$\alpha$ flux inferred from the $K_{\rm s}$-band flux excess than the spectroscopic measurement which may be attributed to their extended star formation distributions.
The low-mass galaxies exhibit extreme [\text{O}\,\textsc{iii}] $\lambda 5007$ emission line with high rest-frame equivalent widths ($\text{EW}_{\text{rest}}$) ranging from 480 to $1064$ \AA. They are characterized by high [\text{O}\,\textsc{iii}]/H$\beta$ ratios ($>4$) and weak [\text{O}\,\textsc{ii}] and [\text{N}\,\textsc{ii}] emission lines, indicating high ionization parameters and low metallicity. The line ratios are comparable to those of star-forming galaxies at $z \gtrsim 6$ and their local analogs.
JWST/NIRCam imaging reveals a clear mass-dependent color gradient: the low-mass galaxies present positive F115W-F277W radial gradients with blue cores, while the high-mass galaxies exhibit negative gradients with red cores. These results suggest distinct assembly pathways for low- and high-mass galaxies during the cosmic noon epoch.
\end{abstract}



\section{Introduction}
The epoch around redshift $z\sim2$, corresponding to a lookback time of approximately 10 billion years, marks a pivotal period in the cosmic history known as cosmic noon. During this era, the cosmic star formation rate density (SFRD) reached its maximum, with galaxies converting gas into stars at rates more than an order of magnitude higher than observed in the local universe \citep{Madau_2014}. This period is a crucial time for assembly of a substantial fraction of stellar mass found in galaxies today. Understanding the physical mechanisms that drove this elevated level of star formation is therefore essential for a comprehensive picture of galaxy formation and evolution. Theoretical models suggest that elevated star formation during cosmic noon was driven by efficient cold-gas accretion from the cosmic web \citep{Kere__2005,Dekel_2009} and regulated by stellar and AGN feedback, particularly in low-mass galaxies where shallow gravitational potentials enhance outflows \citep{Hopkins_2014, Somerville_2015, Faucher_Gigu_re_2017}. Disentangling relative importance of these processes observationally remains challenging, especially at high redshift and low stellar mass.

Vast majority of star-forming galaxies at $z \sim 2$ exhibit a strong positive correlation between their stellar mass ($M_\star$) and star formation rate (SFR), commonly referred to as the Star-Forming Main Sequence (SFMS). This correlation, observed from the local universe out to high redshifts, suggests that for most galaxies star formation proceeds as a relatively steady process regulated primarily by stellar mass \citep{Daddi_2007,Rodighiero_2011,Speagle_2014}. While the SFMS evolves with redshift, indicating higher gas fractions and enhanced star formation activity in the early universe, a significant scatter exists around this relation. Possible causes of the scatter include differences in gas fraction at fixed stellar mass, variations in gas accretion histories, and environmental effects \citep{Dutton_2010, Matthee_2019}. Short-timescale variability driven by stochastic gas accretion and feedback may contribute substantially to this dispersion, particularly in low-mass galaxies \citep{Tacchella_2016}. Galaxies lying significantly above the SFMS, with high specific star formation rates (sSFRs), are therefore of particular interest as potential sites of rapid or episodic stellar mass assembly.

To effectively study the galaxy population at $z \sim 2$, robust methods for identifying star-forming galaxies and quantifying their SFRs are essential. The H$\alpha$ emission line, originating from ionized gas in H\,\textsc{ii} regions surrounding young massive stars, provides a well-established tracer of recent star formation. At $z = 2.1$--$2.5$, H$\alpha$ falls within the $K_{\rm s}$-band and can be identified through a $K_{\rm s}$-band flux excess relative to the stellar continuum estimated from SED fitting. Following  this method, \citet{Terao_2022} and \citet{Chen_2024} identified H$\alpha$ emitters at $z \sim 2$, including galaxies with stellar masses below \(10^9\,M_\odot\), enabling population analyses and spectroscopic follow-up research.

Despite their importance, comprehensive spectroscopic characterizations of low-mass galaxies ($M_\star < 10^{9}\,M_\odot$) with high sSFRs at $z \sim 2$ remain relatively sparse. Near-infrared spectroscopic surveys such as MOSDEF have made significant progress, but their magnitude-limited target selection (e.g., $H_{\rm AB} < 24.5$ at $z \sim 2.3$) results in samples that predominantly probe stellar masses above $\log(M_\star/M_\odot) \sim 9$ \citep{Kriek_2015}. Similarly, the ZFIRE survey contains only a few low-mass HAEs at $z \sim 2$ \citep{2016ApJ...828...21N}. Low-mass with high-sSFR galaxies have also been studied as Lyman-$\alpha$ emitters at similar redshifts, but such samples often lack direct H$\alpha$ measurements \citep{Vargas_2014,Kusakabe_2015}, complicating direct comparisons with SFMS-based frameworks. The NIR grism survey by HST (3D-HST) has identified numerous low-mass star-forming galaxy at $z \sim 2$, but the H$\alpha$ emission line falls outside the wavelength coverage of its grism spectroscopy, precluding direct measurements of the primary star-formation tracer. As a result, the physical conditions of low-mass, high-sSFR galaxies at $z \sim 2$ remain relatively poorly constrained.

In this paper, we present spectroscopic observations of seven HAE candidates at $z \sim 2$ identified by \citet{Chen_2024} in the ZFOURGE--COSMOS field. Our primary sample consists of low-mass galaxies ($M_\star < 10^{9}\,M_\odot$) exhibiting high sSFRs. For comparison, we also include a reference sample of more massive galaxies ($M_\star > 10^{10}\,M_\odot$) located on the SFMS, enabling comparison with typical SFMS galaxies at the same epoch.

The objectives of this study are to spectroscopically confirm the nature of our candidates and assess the effectiveness of the $K_{\rm s}$-band flux excess method to identify low-mass HAEs at $z \sim 2$. We also constrain their interstellar medium properties, including gas-phase metallicity and ionization conditions, and aim to gain insight into the processes associated with enhanced star formation and stellar mass growth in low-mass galaxies during cosmic noon.

The outline of this paper is as follows. In Section~2, we describe sample selection criteria, observation, and data reduction procedures. In Section~3, we outline the methods used to derive key physical properties, including redshifts, star formation rates, and emission-line measurements. Section~4 presents our observational results, including the distribution of our targets on ionization diagnostic diagrams. In Section~5, we discuss these results in the context of stellar mass and galaxy structure by incorporating JWST/NIRCam imaging. Finally, our conclusions are summarized in Section~6. Throughout this work, we adopt a \citet{Chabrier_2003} initial mass function and assume a flat $\Lambda$CDM cosmology with $H_{0} = 70~\mathrm{km,s^{-1},Mpc^{-1}}$, $\Omega_{\mathrm{m}} = 0.3$, and $\Omega_{\Lambda} = 0.7$.

\begin{figure}[htbp]
\centering
\includegraphics[width=\linewidth]{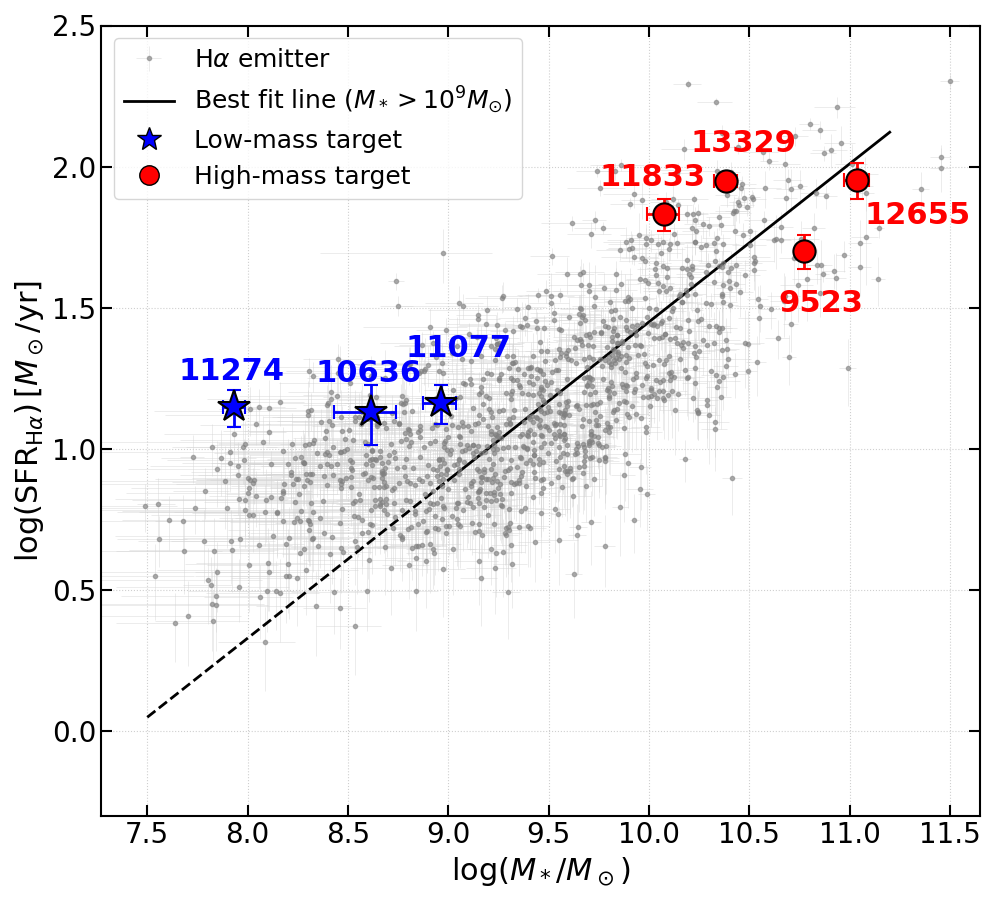}
\caption{
Distribution of HAEs and target galaxies in the ZFOURGE--COSMOS field.
Gray data points denote the parent sample of HAEs at $z \sim 2$.
The black solid line shows the best-fit star-forming main sequence derived from galaxies with $M_\ast > 10^9\,M_\odot$,
while the black dashed line represents its extrapolation toward the lower-mass regime.
}
\label{fig:ms_relation}
\end{figure}

\section{Observation and Data Reduction} \label{sec:style}

\subsection{Target Selection}

We selected seven H$\alpha$ emitter candidates at $z \sim 2$ in the ZFOURGE-COSMOS field from the sample of \citet{Chen_2024}. The parent sample was identified based on a $K_{\mathrm{s}}$-band flux excess relative to the best-fit stellar continuum derived from SED fitting. The sample is listed in Table~\ref{tab:zfourge_properties} and shown as gray points in Figure~\ref{fig:ms_relation}.

Three low-mass HAE galaxies are selected to fulfill the following three criteria; (1) to have a stellar mass $M_* < 10^9 \, M_\odot$, (2) to have sSFR exceeding twice that of the main sequence, within the measurement uncertainties, and (3) to have $F_{\text{H}\alpha} > 4 \times 10^{-17} \, \text{erg s}^{-1} \, \text{cm}^{-2}$ which corresponds to an SFR of $> 7 \, M_\odot \, \text{yr}^{-1}$ at $z \sim 2$. Among the selected low-mass galaxies, ID~11077 has an H$\alpha$ emission-line measurement from the MOSDEF survey, while spectroscopic redshifts are available for IDs~10636 and~11077 from the 3D-HST survey.

For comparison, we included four high-mass galaxies ($M_* > 10^{10} \, M_\odot$) located within the same field of view and situated near the star-forming main sequence (SFMS) at $z \sim 2$ (IDs: 9523, 11833, 12655, 13329). The low-mass and high-mass galaxies are marked by blue stars and red circles, respectively, in Figure~\ref{fig:ms_relation}. In addition, two reference stars in the field was also observed.

%%%%%%%%%%%%%%%%%%%%%%%%%%%%%%%%%%%%%%%%%%%%%%%%%%%%%%%%%%%%
% Table 1 - start
\begin{table*}[htbp]
\centering
\caption{Properties of the Target Galaxies Obtained by SED Fitting}
\label{tab:zfourge_properties}
\begin{tabular*}{\textwidth}{@{\extracolsep{\fill}}cccccccc}
\hline
ZFOURGE ID & R.A. (J2000) & Dec. (J2000) & 
$z_{\mathrm{phot}}$\textsuperscript{a} & 
$\log(M_\ast/M_\odot)$ & 
$\log(\mathrm{SFR}_{\mathrm{H}\alpha})$ &
$E(B-V)$\textsuperscript{b} \\
\hline
\multicolumn{7}{c}{\textit{Low-mass galaxies ($\log(M_{\ast}/M_{\odot}) < 9$)}} \\
\hline
10636 & 10:00:27.234 & +02:17:31.53 & 2.3101 & $8.613^{+0.128}_{-0.181}$ & $1.133^{+0.093}_{-0.118}$ & $0.062 \pm 0.018$ \\
11077 & 10:00:28.626 & +02:17:48.62 & 2.1023 & $8.966^{+0.075}_{-0.091}$ & $1.163^{+0.063}_{-0.073}$ & $0.083 \pm 0.012$ \\
11274 & 10:00:28.324 & +02:17:58.31 & 2.4794 & $7.935^{+0.051}_{-0.058}$ & $1.148^{+0.061}_{-0.071}$ & $0.067 \pm 0.012$ \\
\hline
\multicolumn{7}{c}{\textit{High-mass galaxies ($\log(M_{\ast}/M_{\odot}) > 10$)}} \\
\hline
9523$^{\dagger}$  & 10:00:33.907 & +02:16:41.49 & 2.1806 & $10.771^{+0.035}_{-0.039}$ & $1.702^{+0.056}_{-0.064}$ & $0.395 \pm 0.033$ \\
11833 & 10:00:26.074 & +02:18:12.11 & 2.1491 & $10.078^{+0.073}_{-0.088}$ & $1.833^{+0.053}_{-0.061}$ & $0.366 \pm 0.023$ \\
12655 & 10:00:25.295 & +02:18:46.22 & 2.2773 & $11.038^{+0.058}_{-0.067}$ & $1.954^{+0.058}_{-0.067}$ & $0.588 \pm 0.041$ \\
13329 & 10:00:27.044 & +02:19:09.96 & 2.1648 & $10.385^{+0.054}_{-0.061}$ & $1.948^{+0.024}_{-0.026}$ & $0.271 \pm 0.026$ \\
\hline
\end{tabular*}

\vspace{1em} 

\begin{minipage}{\textwidth}
\small
\textbf{Notes.}\\
ZFOURGE IDs, coordinates, and photometric redshifts were obtained from the ZFOURGE catalog (COSMOS v1.3.6; \citealt{Straatman_2016}). 
 \\
\textsuperscript{a} The $z_{\text{phot}}$ values were specifically derived using the FAST code \citep{2009ApJ...700..221K}.
 \\
\textsuperscript{b} $E(B-V)$ values correspond to the stellar continuum attenuation derived from SED fitting.
 \\
$^{\dagger}$ Confirmed as an AGN through X-ray cross-matching with the catalog presented by \citet{Cowley_2016}.

\end{minipage}
\end{table*}
% Table 1  - End
%%%%%%%%%%%%%%%%%%%%%%%%%%%%%%%%%%%%%%%%%%%%%%%%%%%%%%%%%%%%

\subsection{Spectroscopic Observations}

Multi-object spectroscopic observations of the seven galaxies were conducted on February 5--6, 2022, using SWIMS on the Subaru Telescope (Proposal ID: S22A0045N).
SWIMS provides continuous wavelength coverage from 0.9--1.4~$\mu$m in the blue arm and 1.4--2.5~$\mu$m in the red arm \citep{2020SPIE11447E..5HK}. This enables us to simultaneously capture major rest-frame optical emission lines, including [O~\textsc{ii}]~$\lambda3727$, H$\beta$, 
[O~\textsc{iii}]~$\lambda\lambda4959,5007$, H$\alpha$, [N~\textsc{ii}]~$\lambda\lambda6548,6584$, and
[S~\textsc{ii}]~$\lambda\lambda6716,6731$.
We used a multi-object slit mask with slits of $0\farcs5$ in width and $10\arcsec$ in length, which deliver a spectral resolution of $R \sim 1000$. Observations were carried out with an ABAB nodding sequence. The individual exposure time was 300 seconds to facilitate accurate sky subtraction, and the total on-source exposure time was approximately 250 minutes.

\subsection{Data Reduction}

The data reduction was performed following procedures adopted for near-infrared multi-object spectroscopy, primarily based on the reduction procedure developed for Subaru/MOIRCS. First, flat-fielding, bad-pixel masking, and detector mosaicking were applied to each raw frame. Individual spectral regions corresponding to each slit were extracted prior to wavelength calibration. Wavelength calibration was performed using night-sky OH emission lines, following the wavelengths of the OH lines reported by \cite{2000A&A...354.1134R}. Sky subtraction was performed using an A–B nodding sequence to remove the dominant OH airglow emission. Residual background features were further removed through additional sky subtraction, and all exposures were coadded to maximize the signal-to-noise ratio, yielding the final two-dimensional spectra for each galaxy.

Spectral response and throughput variations were corrected using observations of spectrophotometric standard stars obtained with the same instrumental configuration. Flux calibration was performed using the spectra of the standard stars. Slit-loss was corrected by comparing the observed fluxes of the reference stars within the mask to their known broadband magnitudes.

\section{Spectral Analysis and Physical Measurements} \label{sec:style}

\subsection{1D and 2D Spectra} \label{sec:spectra}

\begin{figure*}[t]
    \centering

    \includegraphics[height=0.75\textheight, keepaspectratio]{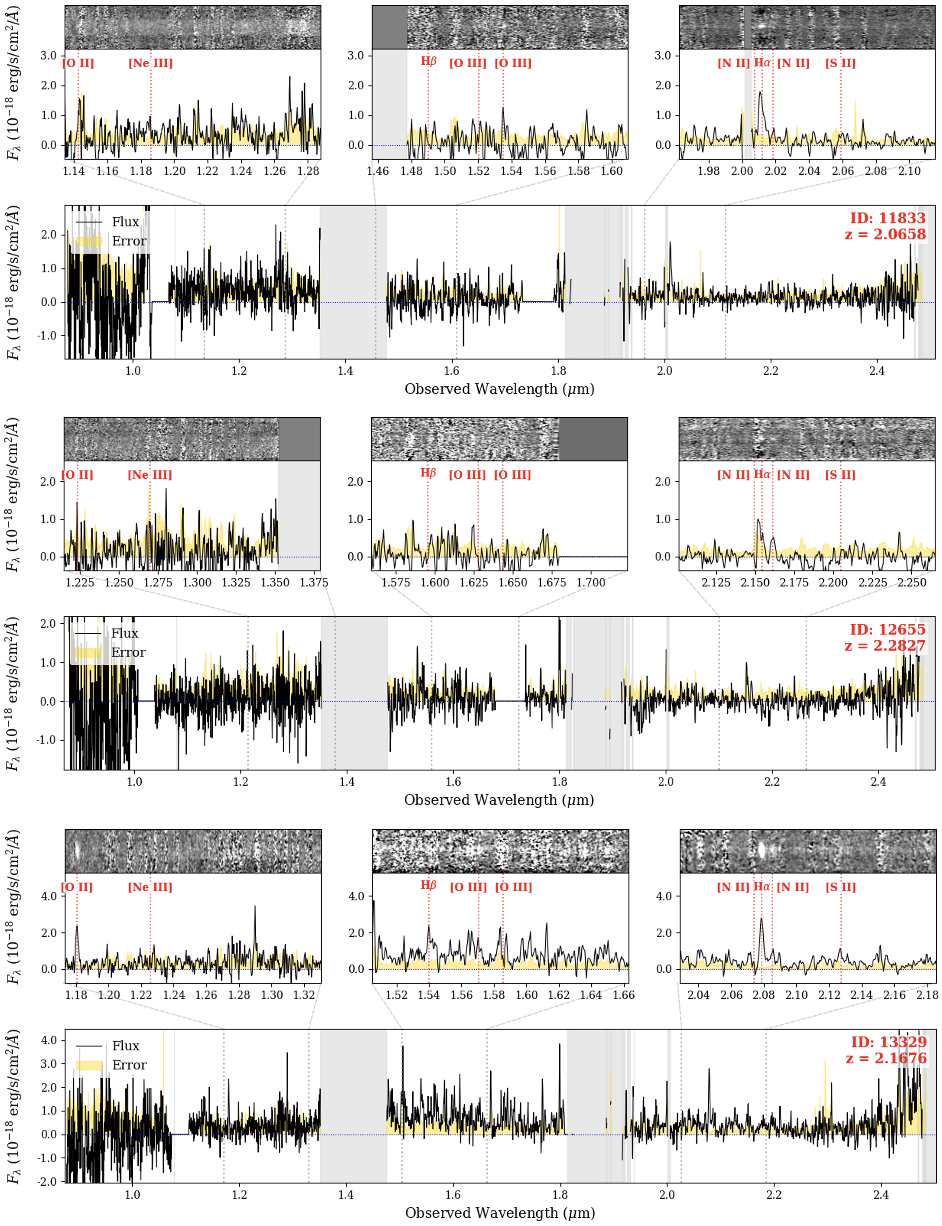} 
    \caption{1D and 2D spectra of the target galaxies. The ID and redshift labels are shown in blue and red for low- and high-mass galaxies, respectively. For each object, the main panel shows the full 1D observed-frame spectrum. The black line represents the flux density, the yellow line denotes the uncertainty, and grey-shaded areas indicate masked regions due to low response function or detector gaps. Above each main plot, three zoomed-in panels highlight key emission-line complexes: [O\,{\sc ii}] $\lambda3727$ (left); H$\beta$ and [O\,{\sc iii}] $\lambda\lambda4959, 5007$ (center); and H$\alpha$, 
    [N\,{\sc ii}] $\lambda\lambda6548, 6584$, and [S\,{\sc ii}] $\lambda\lambda6717, 6731$ (right).}
    \label{fig:all_specs}
\end{figure*}

The 1D and 2D spectra of the seven galaxies are presented in Figure~\ref{fig:all_specs}. We successfully detected the H$\alpha$ emission line in all seven galaxies, yielding secure spectroscopic redshifts within the range of $2.06 < z < 2.49$. The low-mass and high-mass galaxies exhibit distinct spectroscopic characteristics.

The low-mass galaxies (IDs 10636, 11077, and 11274) show prominent [O\,{\sc iii}] emission line, where both components of the [O\,{\sc iii}]\,$\lambda\lambda4959,5007$ doublet are clearly detected with high signal-to-noise ratios, and the [O\,{\sc iii}]\,$\lambda5007$ emission line is stronger than H$\alpha$. In contrast, [N\,{\sc ii}] and [O\,{\sc ii}] emission lines are not detected.

Unlike the low-mass galaxies, three of the four high-mass galaxies (IDs 9523, 11833, 12655, and 13329) show detectable [N\,{\sc ii}] emission line. Furthermore, the overall [O\,{\sc iii}]\,$\lambda5007$/H$\alpha$ ratios are lower than those observed in the low-mass galaxies.

\subsection{Emission Line Measurement}
We measured the emission line fluxes by fitting Gaussian profiles using the \texttt{lmfit} package \citep{2016ascl.soft06014N}. Following the methodology of \citet{2025ApJ...982...27Z}, each line was modeled as a Gaussian component combined with a constant continuum baseline. For blended lines, multiple Gaussian components were fitted simultaneously to ensure accurate deblending.

The total line flux was determined by integrating the continuum-subtracted spectrum over a wavelength range of $\pm 3\sigma_{\rm line}$ centered on the line centroid, where $\sigma_{\rm line}$ is the standard deviation of the best-fitting Gaussian profile. The resulting line fluxes are listed in Table~\ref{tab:zfourge_lines}. For emission lines that were not detected, we derived $3\sigma$ upper limits on their fluxes, where $\sigma$ represents the measured flux uncertainty.

\subsection{Redshift Determination}

Spectroscopic redshifts were determined using the H$\alpha$ emission line detected in the reduced 1D and 2D spectra. All the galaxies lie within the redshift range of $z \sim 2.0$--2.5, as expected from their photometric redshifts.

The redshifts independently derived from the other emission lines are in excellent agreement with $|\Delta z|/(1+z) \lesssim 1\times10^{-5}$ for all the galaxies, confirming the robustness of the redshift determination. We therefore adopt the H$\alpha$-based spectroscopic redshifts as the final redshift values used throughout this work.

The obtained redshifts are consistent with available spectroscopic redshifts from previous studies, including MOSDEF \citep{Kriek_2015}, 3D-HST \citep{2016ApJS..225...27M, 2012ApJS..200...13B}, and ZFIRE \citep{2016ApJ...828...21N} within the measurement uncertainties.

For ID 11274, only a photometric redshift had previously been available. Our Subaru/SWIMS observations detected the H$\alpha$ emission line, providing the first spectroscopic confirmation of this galaxy at $z = 2.4908$.

\subsection{Equivalent Width}

The rest-frame equivalent widths ($\mathrm{EW}_{\mathrm{rest}}$) of the [O\,{\sc iii}]\,$\lambda5007$ emission lines were determined by dividing the measured line fluxes by the underlying stellar continuum levels. These continuum levels were estimated from best-fit SED models derived using multi-wavelength photometric data spanning 0.3--8\,$\mu$m from the ZFOURGE catalog combined with SWIMS photometry. 

The resulting $\mathrm{EW}_{\mathrm{rest}}$ values are $556.30 \pm 69.21\,\text{\AA}$, $480.42 \pm 29.37\,\text{\AA}$, and $1064.05 \pm 271.68\,\text{\AA}$ for IDs 10636, 11077, and 11274, respectively. These values classify the low-mass galaxies as Extreme Emission Line Galaxies, which are defined as galaxies with $\mathrm{EW}_{\mathrm{rest}} > 300\,\text{\AA}$ \citep{del_Moral_Castro_2024}.

%%%%%%% 1D & 2D spectrum - start %%%%%%%
% --- 1st page (ID 9523 ~ 11274, 4개 행) ---
\begin{figure*}[p]
    \centering

    \includegraphics[height=1.00\textheight, keepaspectratio]{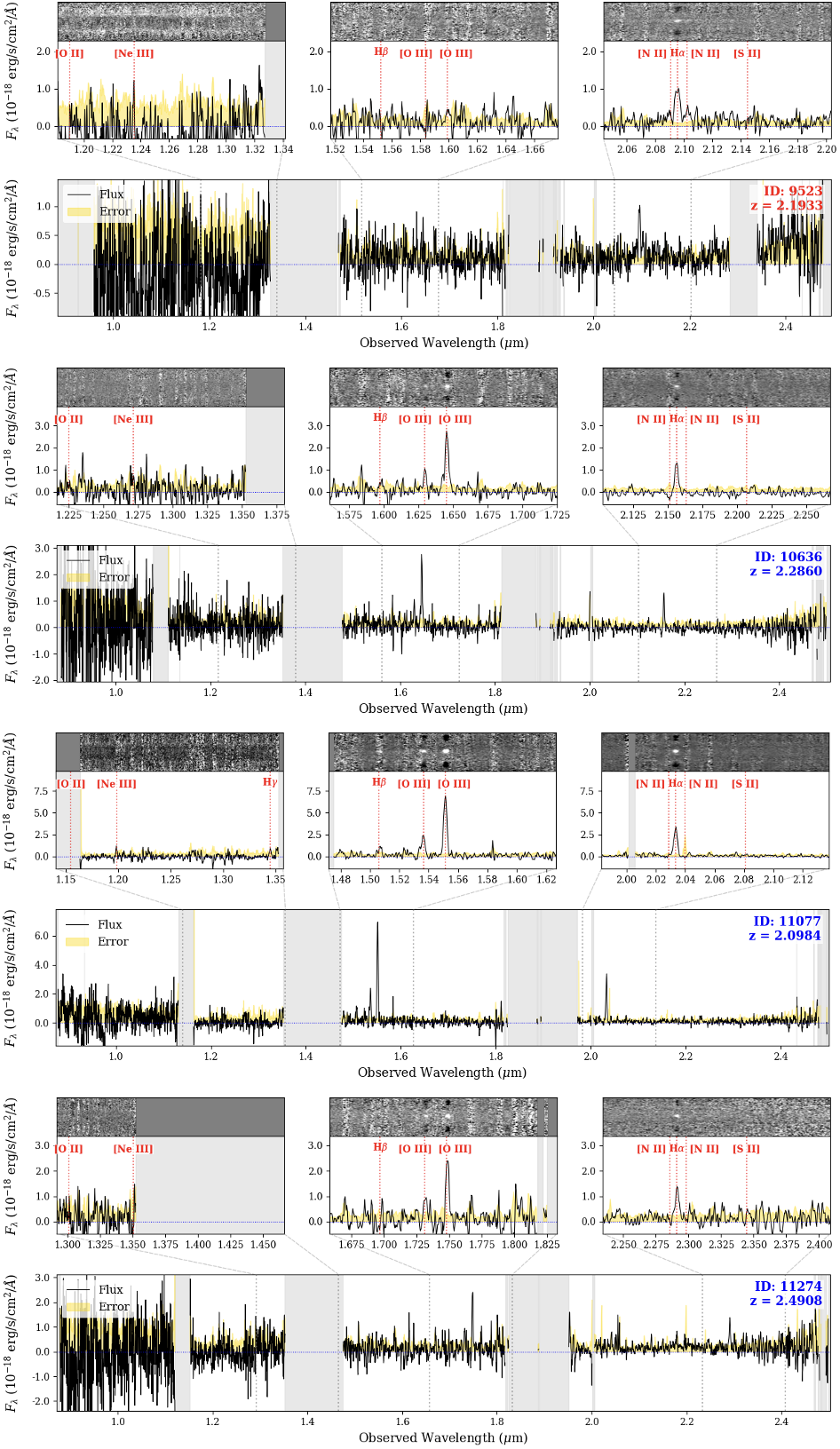} 
\end{figure*}

% --- 2nd page (ID 11833 ~ 13329, 3개 행) ---
%%%%%%% 1D & 2D spectrum - end %%%%%%%

%%%%%%%%%%%%%%%%%%%%%%%%%%%%%%%%%%%%%%%%%%%%%%%%%%%%%%%%%%%%
% Table 2 - start -------------------------------------------
\begin{table*}[t]
\centering

\caption{Emission-Line Measurements of Target Galaxies from Subaru/SWIMS Spectroscopic Observations}
\label{tab:zfourge_lines}
% \vspace{0.5em}

\small
\renewcommand{\arraystretch}{0.95}

\begin{tabular*}{\textwidth}{@{\extracolsep{\fill}}c c l c} % 열 개수를 5개에서 4개로 수정
\hline   % 표의 맨 윗부분 두 줄
ZFOURGE ID & $z_{\rm spec}$\textsuperscript{a} & Emission Line & Flux\textsuperscript{b} \\
 & & & ($10^{-18}$ erg s$^{-1}$ cm$^{-2}$) \\

\hline % 헤더 아래 구분선

\multicolumn{4}{c}{\textit{Low-mass galaxies ($\log(M_{\ast}/M_{\odot}) < 9$)}} \\ % multicolumn 5 -> 4
\hline
10636 & 2.2860 & [O\,{\sc ii}] $\lambda\lambda3727,3729$ & $<15.37$ \\
      &        & H$\beta$                             & $<11.33$ \\
      &        & [O\,{\sc iii}] $\lambda4959$          & $20.49 \pm 3.19$ \\
      &        & [O\,{\sc iii}] $\lambda5007$          & $84.76 \pm 5.06$ \\
      &        & H$\alpha$                            & $43.46 \pm 3.31$ \\
      &        & [N\,{\sc ii}] $\lambda6584$           & $<5.80$ \\
\hline % ID 구분선
11077 & 2.0984 & [Ne\,{\sc iii}] $\lambda3869$         & $14.96 \pm 4.26$ \\
      &        & H$\gamma$                            & $19.06 \pm 4.58$ \\
      &        & H$\beta$                             & $22.37 \pm 7.52$ \\
      &        & [O\,{\sc iii}] $\lambda4959$          & $63.98 \pm 5.50$ \\
      &        & [O\,{\sc iii}] $\lambda5007$          & $187.60 \pm 4.78$ \\
      &        & H$\alpha$                            & $91.16 \pm 3.60$ \\
      &        & [N\,{\sc ii}] $\lambda6584$           & $<46.66$ \\
\hline % ID 구분선
11274 & 2.4908 & [O\,{\sc ii}] $\lambda\lambda3727,3729$ & $<24.48$ \\
      &        & H$\beta$                             & $<15.72$ \\
      &        & [O\,{\sc iii}] $\lambda4959$          & $30.93 \pm 5.40$ \\
      &        & [O\,{\sc iii}] $\lambda5007$          & $61.13 \pm 5.57$ \\
      &        & H$\alpha$                            & $39.76 \pm 4.34$ \\
      &        & [N\,{\sc ii}] $\lambda6584$           & $<11.43$ \\

\hline  % 그룹 구분선
\multicolumn{4}{c}{\textit{High-mass galaxies ($\log(M_{\ast}/M_{\odot}) > 10$)}} \\ % multicolumn 5 -> 4
\hline
9523$^{\dagger}$  & 2.1933 & [O\,{\sc ii}] $\lambda\lambda3727,3729$ & $<19.41$ \\
      &        & H$\beta$                             & $<11.89$ \\
      &        & [O\,{\sc iii}] $\lambda5007$          & $7.64 \pm 6.07$ \\
      &        & H$\alpha$                            & $40.48 \pm 2.10$ \\
      &        & [N\,{\sc ii}] $\lambda6584$           & $10.97 \pm 2.35$ \\
\hline % ID 구분선
11833 & 2.0658 & [O\,{\sc ii}] $\lambda\lambda3727,3729$ & $<31.91$ \\
      &        & H$\beta$                             & $<23.16$ \\
      &        & [O\,{\sc iii}] $\lambda5007$          & $10.71 \pm 4.32$ \\
      &        & H$\alpha$                            & $59.22 \pm 6.35$ \\
      &        & [N\,{\sc ii}] $\lambda6584$           & $<13.99$ \\
\hline % ID 구분선
12655 & 2.2827 & [O\,{\sc ii}] $\lambda\lambda3727,3729$ & $<16.28$ \\
      &        & H$\beta$                             & $<11.58$ \\
      &        & [O\,{\sc iii}] $\lambda4959$          & $<8.32$ \\
      &        & [O\,{\sc iii}] $\lambda5007$          & $<12.17$ \\
      &        & H$\alpha$                            & $24.25 \pm 4.38$ \\
      &        & [N\,{\sc ii}] $\lambda6584$           & $11.61 \pm 2.21$ \\
\hline % ID 구분선
13329 & 2.1676 & [O\,{\sc ii}] $\lambda\lambda3727,3729$ & $39.74 \pm 2.65$ \\
      &        & H$\beta$                             & $<16.66$ \\
      &        & [O\,{\sc iii}] $\lambda4959$          & $<9.94$ \\
      &        & [O\,{\sc iii}] $\lambda5007$          & $<29.62$ \\
      &        & H$\alpha$                            & $61.35 \pm 1.69$ \\
      &        & [N\,{\sc ii}] $\lambda6584$           & $13.87 \pm 2.81$ \\

\hline % 표의 맨 아랫부분 두 줄
\end{tabular*}

\vspace{0.8em}

\begin{flushleft}
\textbf{Notes.} \\
$^{a}$ Redshifts measured via H$\alpha$ emission lines from this study. \\
$^{b}$ Upper limits (``$<$'') correspond to $3\sigma$ limits. \\
$^{\dagger}$ Confirmed as an AGN through X-ray cross-matching with the catalog presented by \citet{Cowley_2016}.
\end{flushleft}

\end{table*}
% Table 2 - end -------------------------------------------
%%%%%%%%%%%%%%%%%%%%%%%%%%%%%%%%%%%%%%%%%%%%%%%%%%%%%%%%%%%%

%%%%%%%%%%%%%%%%%%%%%%%%%%%%%%%%%%%%%%%%%%%%%%%%%%%%%%%%%%%%
% Table 3 - Line ratio - Start
\begin{table*}[t] 
\centering
\caption{Emission-Line Ratios for the BPT, $O32$--$R23$, and $R3$--$R2$ Diagrams}
\label{tab:line_ratios_3decimal}
\smallskip

\begin{tabular*}{\textwidth}{@{\extracolsep{\fill}}cccccc} 
\hline
ZFOURGE ID & $R3$ & $N2$ & $O32$ & $R23$ & $R2$ \\
\hline
\multicolumn{6}{c}{\textit{Low-mass galaxies ($\log(M_{\ast}/M_{\odot}) < 9$)}} \\
\hline
10636 & $5.951 \pm 0.589$ & $< 0.133$ & $6.312 \pm 3.931^{H}$ & $9.602 \pm 4.309^{H}$ & $1.018 \pm 0.776^{H}$ \\
11077 & $8.301 \pm 2.799$ & $0.031 \pm 0.009^{M}$ & $10.180 \pm 3.149^{M}$ & $8.877 \pm 0.903^{M}$ & $0.669 \pm 0.212^{M}$ \\
11274 & $4.716 \pm 0.673$ & $< 0.287$ & $> 2.291$ & $< 9.164$ & $< 2.058$ \\
\hline
\multicolumn{6}{c}{\textit{High-mass galaxies ($\log(M_{\ast}/M_{\odot}) > 10$)}} \\
\hline
% 9523  & $0.815 \pm 0.650$ & $0.270 \pm 0.059$ & $0.324 \pm 0.200^{H}$ & $4.307 \pm 2.715^{H}$ & $3.007 \pm 2.128$ \\
9523$^{\dagger}$  & $0.815 \pm 0.650$ & $0.270 \pm 0.059$ & $0.324 \pm 0.200^{H}$ & $4.307 \pm 2.715^{H}$ & $3.007 \pm 2.128$ \\
11833 & $0.758 \pm 0.317$ & $0.194 \pm 0.033^{Z}$ & $> 0.210$ & $< 4.629$ & $< 3.616$ \\
12655 & $< 2.656$ & $0.475 \pm 0.125$ & \dots & $< 12.051$ & $< 7.557$ \\
13329 & $1.070 \pm 0.124^{M}$ & $0.225 \pm 0.046$ & $0.215 \pm 0.028^{M}$ & $6.275 \pm 3.036^{M}$ & $3.480 \pm 0.677$ \\
\hline
\end{tabular*}

\vspace{5pt} 

\begin{flushleft}
\small 
\textbf{Notes.} Superscripts denote the data source: $^{H}$ 3D-HST, $^{M}$ MOSDEF, and $^{Z}$ ZFIRE, while $^{\dagger}$ indicates that ZFOURGE ID 9523 is confirmed as an AGN through X-ray cross-matching with the catalog presented by \citet{Cowley_2016}. Values without superscripts are derived from the Subaru/SWIMS observation. 
All emission-line ratios listed in this table are corrected for dust extinction based on \citet{2000ApJ...533..682C}. For Subaru/SWIMS data, $[\text{O\,\textsc{iii}}]\,\lambda 4959$ fluxes are estimated from $[\text{O\,\textsc{iii}}]\,\lambda 5007$ fluxes by adopting the theoretical ratio $1 : 2.98$ if $[\text{O\,\textsc{iii}}]\,\lambda 4959$ fluxes are not detected.

\end{flushleft}
\end{table*}
% Table 3 - Line ratio  - End
%%%%%%%%%%%%%%%%%%%%%%%%%%%%%%%%%%%%%%%%%%%%%%%%%%%%%%%%%%%%

\section{Results} \label{sec:style}

\subsection{Comparison of H$\alpha$ Fluxes}

\begin{figure}[htbp] 
    \centering

    \includegraphics[width=\columnwidth]{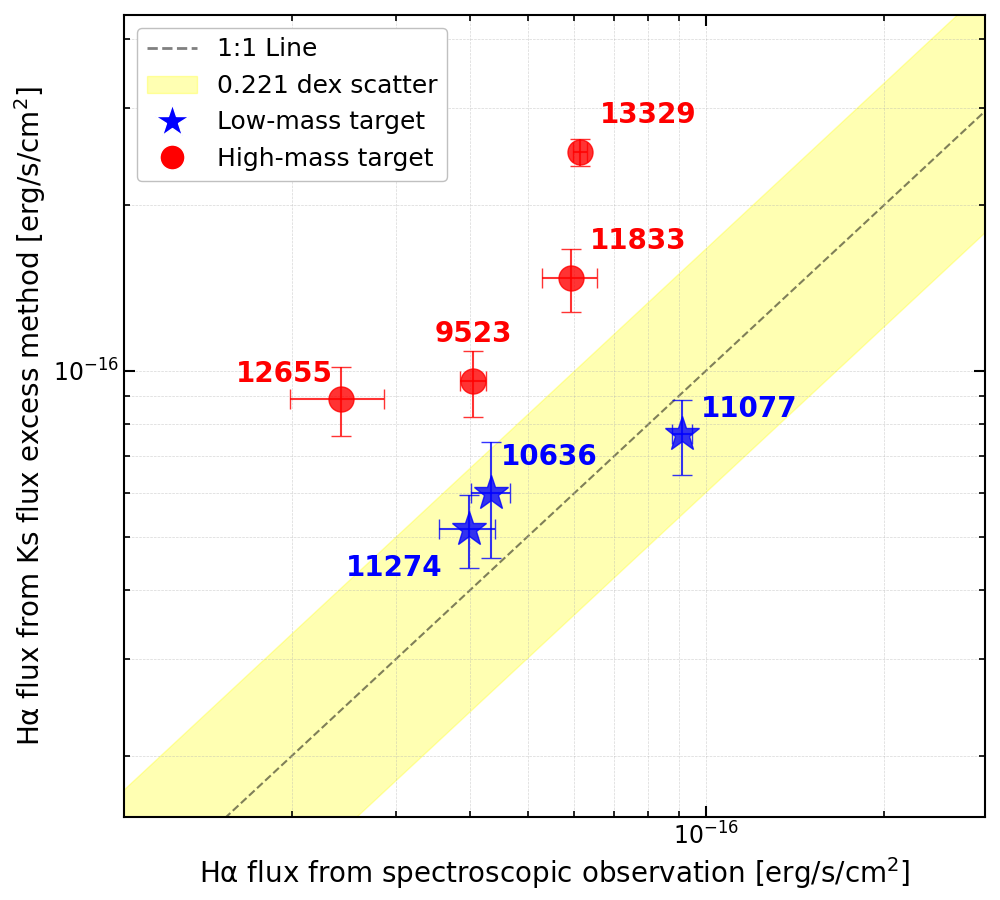} 
    \caption{Comparison of H$\alpha$ fluxes derived from the $K_s$-band flux excess method (y-axis) and Subaru/SWIMS spectroscopic observations (x-axis). Blue stars represent the low-mass galaxies, and red circles represent high-mass galaxies. The dashed line indicates the 1:1 relation, and the yellow shaded region represents the 0.221 dex scatter adopted from \citet{Chen_2024}.}
    \label{fig:ha_flux_comparison}
\end{figure}

We compare the H$\alpha$ fluxes derived from the $K_s$-band flux excess method with our results in Figure~\ref{fig:ha_flux_comparison}. The two methods show varying degrees of agreement depending on the stellar mass of the galaxies.

For the low-mass galaxies, the measured fluxes are in good agreement, with a dispersion smaller than the $0.221\text{ dex}$ scatter (yellow shaded region), which represents the $1\sigma$ dispersion in the differences between the $K_s$-excess and MOSDEF fluxes reported by \citet{Chen_2024}. Although the sample size is limited, this result indicates that the spectroscopic H$\alpha$ fluxes are consistent with those derived from the flux-excess method for low-mass galaxies.

In contrast, the high-mass galaxies tend to show higher H$\alpha$ fluxes in the flux excess method compared to those from spectroscopy. One plausible explanation for this offset is slit loss in the spectroscopic observations. While the low-mass galaxies tend to have compact star-formation distributions that are well-captured within the slit, the high-mass galaxies often possess a more extended H$\alpha$ emission-line distribution \citep{2016ApJ...828...27N, 2019PASJ...71...69S, Terao_2022}. Consequently, a significant fraction of the H$\alpha$ fluxes in the high-mass galaxies may fall outside the slit, leading to the underestimation of the spectroscopic fluxes.

%%%%%%%%%%%%%%%%%%%%%%%%%%%%%%%%%%%%%%%%%%%%%%%%%%
% Figure3 - Ha Flux comparison - start
% Figure3 - Ha Flux comparison - End
%%%%%%%%%%%%%%%%%%%%%%%%%%%%%%%%%%%%%%%%%%%%%%%%%%

%%%%%%%%%%%%%%%%%%%%%%%%%%%%%%%%%%%%%%%%%%%%%%%%%%
% Figure4 - slit-loss - start
\begin{figure*}[t]
    \centering
    \includegraphics[width=1.0\textwidth]{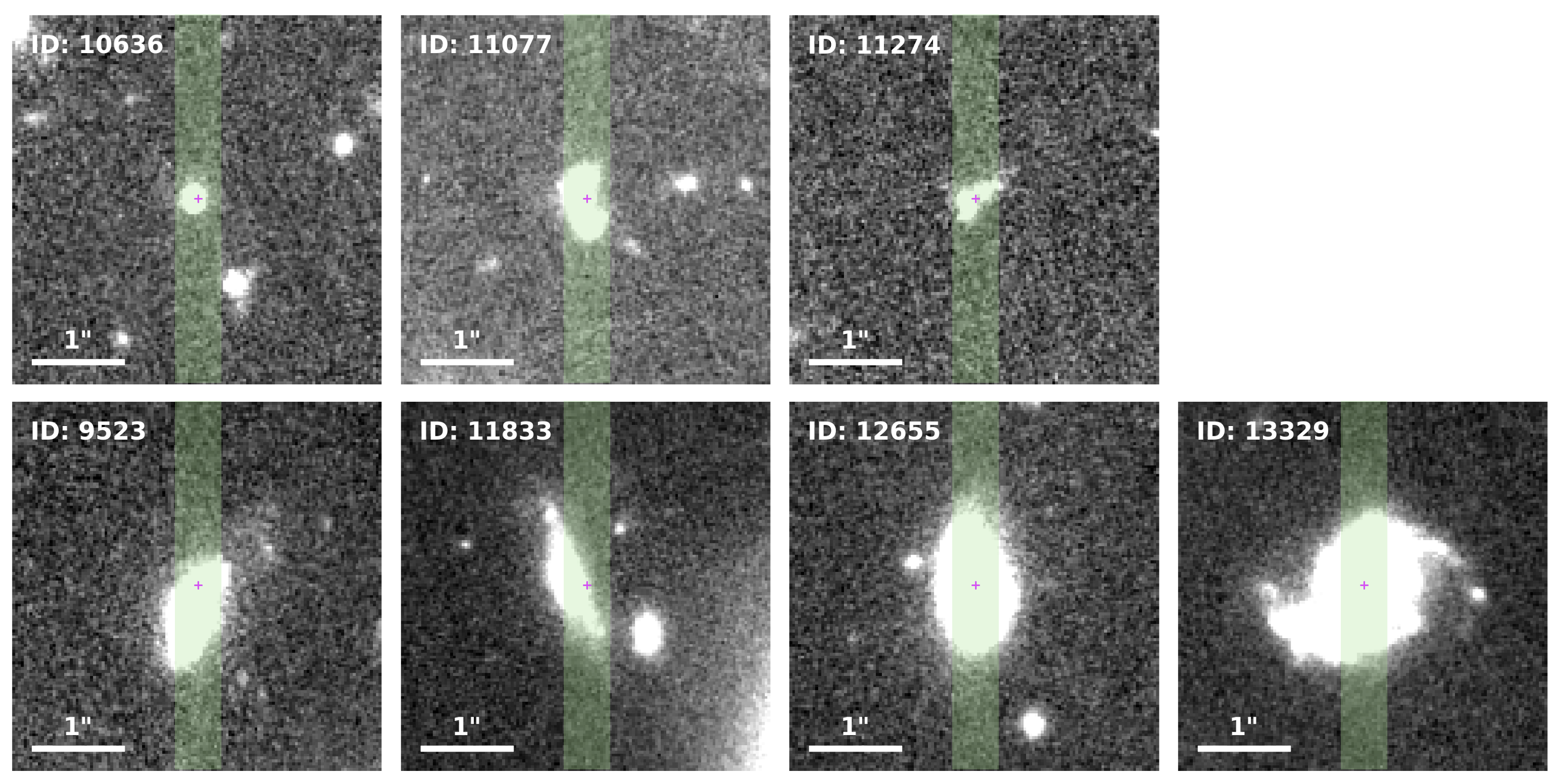}
    \caption{F200W images of the target galaxies. The top row shows the low-mass galaxy (IDs 10636, 11077, 11274), while the bottom row shows the high-mass galaxy (IDs 9523, 11833, 12655, 13329). Each $4'' \times 4''$ FoV panel shows the galaxy morphology with the spectroscopic slit ($0.5'' \times 10''$) position indicated by the green shaded region. The cross marks at the center of images denote the R.A. and Decl. coordinates from the ZFOURGE catalog.}
    \label{fig:target_stamps}
\end{figure*}
\vspace{13pt}
% Figure4 - slit-loss - end
%%%%%%%%%%%%%%%%%%%%%%%%%%%%%%%%%%%%%%%%%%%%%%%%%%

\subsection{Galaxy Morphology and Slit Losses}

To examine the morphological differences between the low- and high-mass galaxies and their potential impact on the spectroscopic measurements, we utilized deep JWST/NIRCam F200W imaging from the JELS survey \citep{2025MNRAS.541.1329D, 2025MNRAS.541.1348P}, as shown in Figure~\ref{fig:target_stamps}. These data are available through the
Mikulski Archive for Space Telescopes (MAST) at
\dataset[doi:10.17909/a97b-3a20]{https://doi.org/doi:10.17909/a97b-3a20}. The F200W imaging primarily traces the stellar continuum and may not exactly represent the spatial distribution of H$\alpha$ emission, which traces recent star formation and can exhibit a different spatial extent from the stellar continuum \citep[e.g.,][]{2024A&A...690A..64M}. Thus, the F200W morphology is used here as a reference for the overall stellar light distribution rather than as a direct tracer of the H$\alpha$-emitting regions. Note that the actual slit alignment during observations may have small pointing offsets relative to the catalog coordinates.

The F200W images reveal a clear distinction in galaxy structures. The low-mass galaxies exhibit compact morphologies, with the bulk of their emission contained within the $0.5''$ width of the slit. This compactness suggests that the majority of their H$\alpha$ emission is likely captured within the spectroscopic aperture, resulting in minimal slit loss and a closer agreement between the photometric and spectroscopic H$\alpha$ estimates.

In contrast, the high-mass galaxies show significantly more extended and diffuse structures. In particular, IDs 12655 and 13329 appear as large systems with structures spreading well beyond the slit boundaries. Such extended morphologies indicate that a non-negligible fraction of the total H$\alpha$ emission likely originates from regions outside the $0.5''$ slit.

Although other factors—such as uncertainties in continuum modeling, spatial variations in
dust attenuation, or astrometric offsets—may also contribute, the morphological distribution suggests that slit loss is a dominant driver of the observed systematic differences. Consequently, the discrepancy observed in Figure~\ref{fig:ha_flux_comparison}, where the spectroscopic H$\alpha$ fluxes for the high-mass galaxies are lower than those from the $K_s$-band excess method, is primarily attributed to the morphological contrast between the two mass regimes relative to the $0.5''$ slit width.

\subsection{Line Ratios}
\label{subsec:line_ratios}

To investigate physical properties of the galaxies, we calculated several key emission-line ratios, namely $R3$, $N2$, $O32$, $R23$, and $R2$, defined as follows:

\begin{equation}
R3 = \frac{[\text{O\,\textsc{iii}}]\,\lambda 5007}{\mathrm{H}\beta}
\end{equation}

\begin{equation}
N2 = \frac{[\text{N\,\textsc{ii}}]\,\lambda 6584}{\mathrm{H}\alpha}
\end{equation}

\begin{equation}
O32 = \frac{[\text{O\,\textsc{iii}}]\,\lambda 5007}{[\text{O\,\textsc{ii}}]\,\lambda\lambda 3726, 3729}
\end{equation}

\begin{equation}
R23 = \frac{[\text{O\,\textsc{ii}}]\,\lambda\lambda 3726, 3729 + [\text{O\,\textsc{iii}}]\,\lambda\lambda 4959, 5007}{\text{H}\beta}
\end{equation}

\begin{equation}
R2 = \frac{[\text{O\,\textsc{ii}}]\,\lambda\lambda 3726, 3729}{\mathrm{H}\beta}
\end{equation}

Line ratios involving emission lines that are close in wavelength, such as $R3$ and $N2$, are relatively insensitive to dust extinction. In contrast, $O32$, $R23$, and $R2$ are sensitive to reddening because of the large wavelength separation between the emission lines. We therefore applied dust extinction corrections to all line ratios assuming the \citet{2000ApJ...533..682C} attenuation law. The stellar color excess, $E(B-V)_{\rm star}$, derived from the SED fitting was adopted and converted to nebular color excess assuming $E(B-V)_{\rm neb}=E(B-V)_{\rm star}$, where stellar and nebular color excesses are comparable at high redshift \citep[e.g.,][]{2016A&A...586A..83P}. Specifically for the Subaru/SWIMS data, the observed H$\beta$ flux was used for ID 11077, where it was detected. For the remaining galaxies with H$\beta$ non-detections, H$\beta$ fluxes were inferred by scaling the $\mathrm{H}\alpha$ fluxes assuming Case~B, where $(\mathrm{H}\alpha/\mathrm{H}\beta)_{\mathrm{int}} = 2.86$. We note that the relationship between stellar and nebular reddening remains uncertain at high redshift \citep[e.g.,][]{2013ApJ...777L...8K, 2020ApJ...902..123R}, and this assumption may introduce systematic uncertainties.

To enable a more complete characterization of the emission-line properties of our targets, particularly for lines that were not detected in our SWIMS observations, we supplemented our SWIMS measurements with archival emission-line data from MOSDEF \citep{Kriek_2015}, 3D-HST \citep{2012ApJS..200...13B, 2016ApJS..225...27M}, and ZFIRE \citep{2016ApJ...828...21N} surveys. For 3D-HST data, individual $[\mathrm{O}\,\textsc{iii}]\, \lambda4959$ and $[\mathrm{O}\,\textsc{iii}]\, \lambda5007$ fluxes were derived from the total $[\mathrm{O}\,\textsc{iii}]$ flux assuming a theoretical ratio of $[\mathrm{O}\,\textsc{iii}]\,\lambda4959:[\mathrm{O}\,\textsc{iii}]\,\lambda5007 = 1:2.98$. We confirmed that the $[\mathrm{O}\,\textsc{iii}]\,\lambda4959/\lambda5007$ line ratios from these archival surveys are consistent with our SWIMS measurements within the corresponding uncertainties.

In cases where multiple measurements were available for a given galaxy, we adopted values according to the following priority: SWIMS, MOSDEF, and 3D-HST. This prioritization reflects differences in spectral resolution and data quality among the surveys. For the ZFIRE sample, which does not overlap with MOSDEF or 3D-HST in our dataset, no relative prioritization was necessary.

We note a specific case regarding ID 13329, which was observed by both MOSDEF and 3D-HST. Visual inspection of the 1D 3D-HST spectrum revealed an erroneous line-fitting result that produced a significant redshift discrepancy relative to both the MOSDEF and SWIMS measurements. Because of this unreliable fit, the 3D-HST measurements for ID 13329 were excluded from our analysis, and only the MOSDEF data were used for this galaxy.

The final set of emission-line ratios is summarized in Table~\ref{tab:line_ratios_3decimal}.

\subsection{BPT Diagram}
\label{subsec:bpt}

\begin{figure}[ht!] % 
    \centering

    \includegraphics[width=\columnwidth]{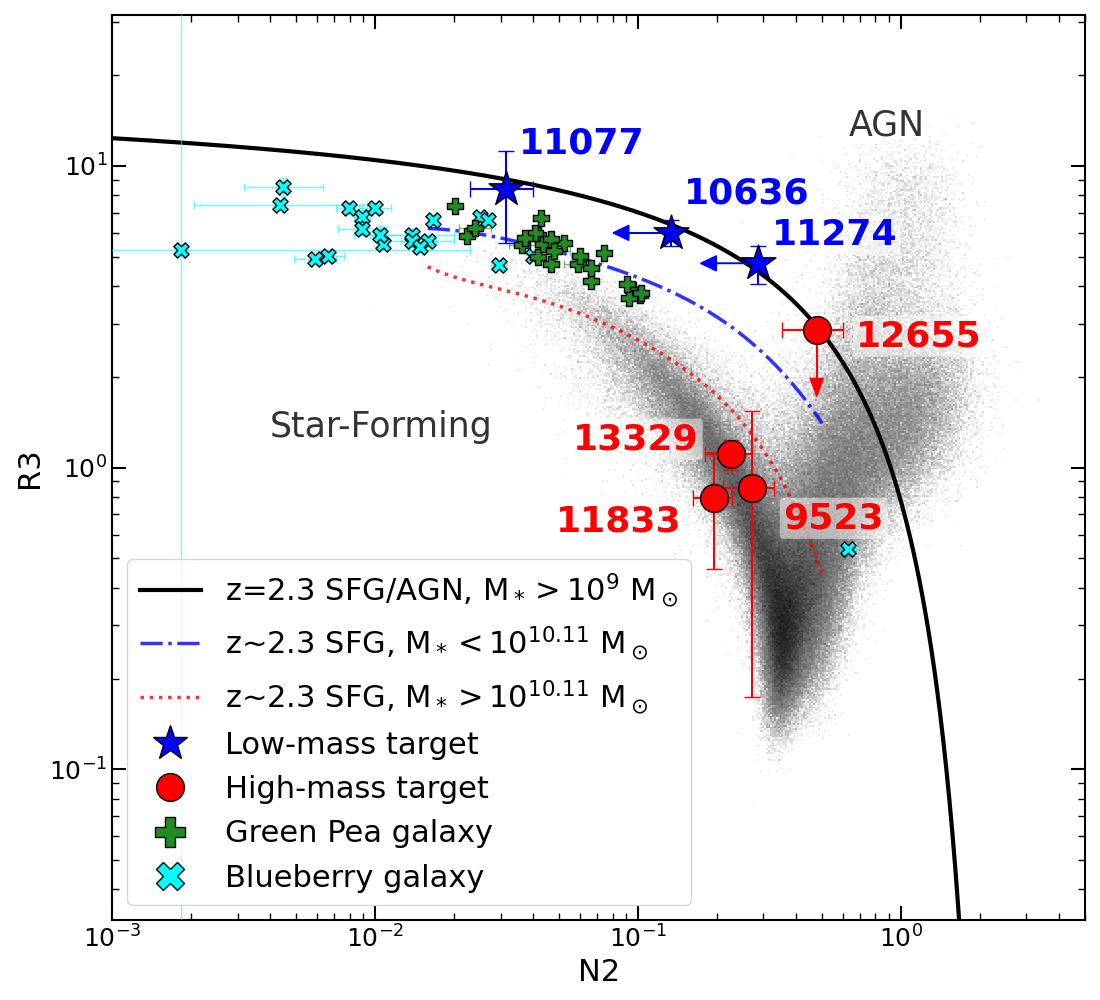} 
    \caption{The BPT diagram for the target galaxies. Blue stars denote the low-mass galaxies, while red circles denote high-mass galaxies. For comparison, Green Pea galaxies \citep{2023A&A...677A.114A} and Blueberry galaxies \citep{2024A&A...688A.159K} are represented by green plus symbols and cyan crosses, respectively. Grayscale dots show SDSS galaxies at $z \sim 0$. Solid black line indicates the star-forming galaxy and AGN classification boundary at $z \sim 2.3$ \citep{2013ApJ...774L..10K}. Blue dash-dotted and red dotted lines represent the fitting of star-forming galaxies at $z \sim 2.3$ with $M_* < 10^{10.11}~M_\odot$ and $M_* > 10^{10.11}~M_\odot$, respectively \citep{2015ApJ...801...88S}.}
    \label{fig:bpt_diagram}
\end{figure}

Figure \ref{fig:bpt_diagram} presents the BPT diagram \citep{1981PASP...93....5B} for the seven galaxies. For comparison, we also plot Green Pea and Blueberry galaxies, which are well-established analogs for intensely star-forming systems at reionization epoch. The low-mass galaxies are located near the $z \sim 2.3$ SFG/AGN boundary, sharing a similar region with the Green Pea and Blueberry galaxies. This tendency is commonly observed in low-mass, intensely star-forming galaxies at high redshift, where extreme ionization conditions drive elevated $R3$ \citep[e.g.,][]{sanders2023excitationionizationpropertiesstarforming}, while low metallicity results in low $N2$. In contrast, the high-mass galaxies exhibit relatively high $N2$, consistent with increased chemical enrichment expected from the mass–metallicity relation \citep{2004ApJ...613..898T,2006ApJ...644..813E}.

The AGN classification for ID~11077 differs depending on the infrared (IR) color--color diagnostic employed. While the criteria of \citet{2012ApJ...754..120M} classify this source as an AGN candidate \citep{Cowley_2016}, the IRAC color selection by \citet{2022MNRAS.509.4940T} places it outside the AGN regime. A recognized limitation of the \citet{2012ApJ...754..120M} selection is its susceptibility to contamination from starburst galaxies, whose warm dust emission can mimic AGN-like red colors. In fact, the IR colors of ID~11077 are closely aligned with the starburst galaxy templates from SWIRE \citep{2007ApJ...663...81P}. In addition, deep X-ray observations show no evidence of nuclear activity, consistent with a normal star-forming galaxy \citep{2014ApJ...790..113Z}. Combining the \citet{2022MNRAS.509.4940T} classification, the vulnerability of the \citet{2012ApJ...754..120M} diagnostic to starburst contamination, and the non-detection in X-rays, we treat ID~11077 as a non-AGN (star-forming galaxy) in this work.

ID 9523, confirmed as an AGN through X-ray cross-matching \citep{Cowley_2016}, falls within the SFG locus. While H$\alpha$ and $[\text{N\,\textsc{ii}}]$ emission lines are robustly detected, the $[\text{O\,\textsc{iii}}]\,\lambda 5007$ emission line is remarkably weak ($S/N \sim 1$). This is consistent with recent findings that X-ray-confirmed AGNs can fall within the SFG locus when their narrow-line AGN component is intrinsically weak \citep{2019ApJ...876...12A}.

Recent studies have also identified AGNs that fall close to the SFG/AGN demarcation line at low $N2$, or that reside within the star-forming locus due to suppressed $R3$, deviating from the classical AGN sequence in the BPT diagram (e.g., \citealt{Scholtz_2025}). These cases highlight the limitations of optical BPT diagnostics for identifying AGNs, and indicate that multi-wavelength indicators are essential for robust classification in high-redshift HAE samples.

%%%%%%%%%%%%%%%%%%%%%%%%%%%%%%%%%%%%%%%%%%%%%%%%%%%%%%
% Figure5 - BPT diagram - start
% Figure5 - BPT diagram - End
%%%%%%%%%%%%%%%%%%%%%%%%%%%%%%%%%%%%%%%%%%%%%%%%%%%%%%

%%%%%%%%%%%%%%%%%%%%%%%%%%%%%%%%%%%%%%%%%%%%%%%%%%%%%%
% Figure6 - O32--R23 & R3--R2 combined diagram - Start
\begin{figure*}[t]
    \centering
    \includegraphics[width=1.0\textwidth]{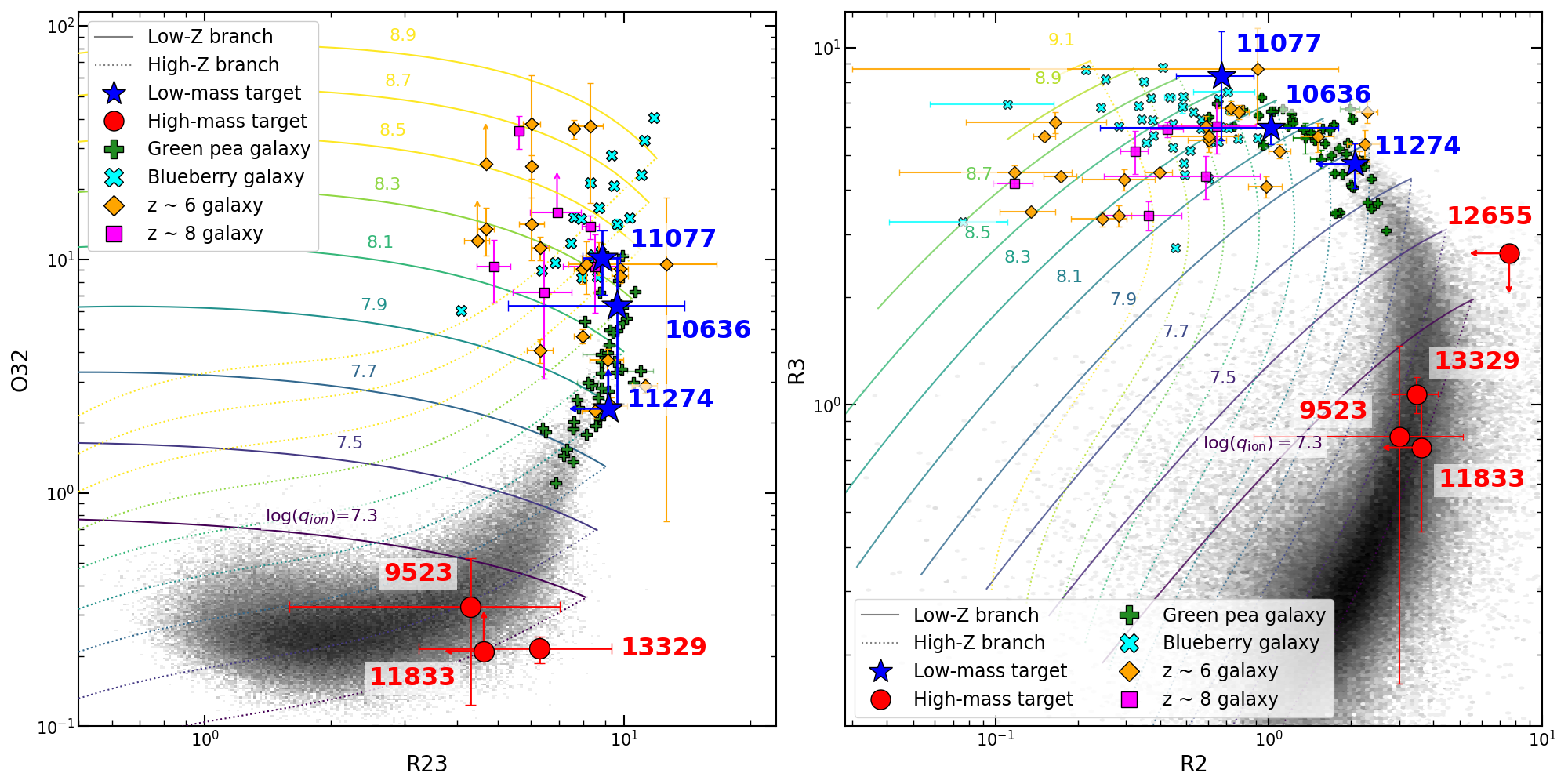}
    \caption{$O32$ versus $R23$ (\textit{left}) and $R3$ versus $R2$ (\textit{right}) diagnostic diagrams including our targets and comparison samples, together with the $z \sim 0$ SDSS galaxy distribution (grayscale density map) and several populations of local and high-redshift analogs. Blue stars represent the low-mass galaxies, and red circles represent high-mass galaxies. Green plusses correspond to Green Pea galaxies from \citet{2017ApJ...844..171Y}, and light blue crosses mark Blueberry galaxies from \citet{2017ApJ...847...38Y}. Orange diamonds and magenta squares show high-redshift galaxies at $z \sim 6$ and $z \sim 8$, respectively, compiled from \citet{2023A&A...677A.115C}. All emission-line fluxes are corrected for dust attenuation using the \citet{2000ApJ...533..682C} reddening law. The colored curves show photoionization model grids based on the calibration of \citet{Kobulnicky_2004}, with solid and dotted lines denoting the low- and high-metallicity branches, respectively, at fixed ionization parameters ($\log (q_{\text{ion}})$) as labeled; the turnover between the two branches is defined at $12+\log(O/H) = 8.4$. In both panels, AGN-contaminated sources have been removed from the SDSS sample following \citet{2003MNRAS.346.1055K}.}
    \label{fig:o32_r23_r3_r2}
\end{figure*}
% Figure6 - O32--R23 & R3--R2 combined diagram - End
%%%%%%%%%%%%%%%%%%%%%%%%%%%%%%%%%%%%%%%%%%%%%%%%%%%%%%

\subsection{O32--R23 and R3--R2 Diagrams}

To investigate the physical conditions of the ionized gas, we compare our galaxies with local and high-redshift star-forming populations using the $O32$--$R23$ and $R3$--$R2$ diagnostic diagrams shown in Figure~\ref{fig:o32_r23_r3_r2}.

In the left panel, the low-mass galaxies are located in the upper-right region of the diagram, characterized by elevated $O32$ and $R23$ near the turnover regime ($R23 \sim 8$--$10$). Their positions overlap with those of the Green Pea and Blueberry galaxies and extend toward the locus of the $z \sim 6$--8 galaxies. This is consistent with the right panel, where the low-mass galaxies show high $R3$ ($R3 > 4$), comparable to that of the Green Pea, Blueberry, and $z > 6$ galaxies; their $R2$ values lie between those of the Green Pea and Blueberry galaxies, and are larger than those of the $z > 6$ galaxies on average. Taken together, these diagnostics indicate that the low-mass galaxies are located near model tracks corresponding to relatively high ionization parameters ($\log q_{\text{ion}} \sim 7.9$--$8.7$). The proximity of these galaxies to the $z \sim 6$--8 galaxy samples in both diagrams reinforces the interpretation that they may serve as $z \sim 2$ analogs of reionization-epoch galaxies, sharing similar physical conditions such as elevated ionization parameters.

In contrast, the distribution of high-mass galaxies overlaps with that of $z \sim 0$ SDSS galaxies. Their low $R3$, high $R2$, and low $O32$ are consistent with a comparatively less intense ionizing radiation field expected for metal-rich star-forming galaxies. The clear separation between the two stellar-mass groups across both diagrams is consistent with the mass--metallicity relation (MZR), in the sense that low-mass galaxies exhibit higher excitation states and lower chemical enrichment compared to the high-mass galaxies.

%%%%%%%%%%%%%%%%%%%%%%%%%%%%%%%%%%%%%%%%%%%%%
%%%%% Table 4 - (metallicity, log q) - Start %%%%%

\begin{table*}[t]
\centering
\small % 
\caption{Metallicity, $\log q$, $R_{e,circ}$, and $\Sigma_{\text{SFR}}$ of Target Galaxies}
\label{tab:galaxy_properties_metal}
\begin{tabular*}{\textwidth}{@{\extracolsep{\fill}}cccccc}
\hline

\makebox[0.14\textwidth][c]{ZFOURGE ID} & 
\makebox[0.14\textwidth][c]{Metallicity ($N2$)$^a$} & 
\makebox[0.14\textwidth][c]{Metallicity ($O3N2$)$^b$} & 
\makebox[0.14\textwidth][c]{$\log q^c$ [cm/s]} & 
\makebox[0.14\textwidth][c]{$R_{e,circ}$ [kpc]$^d$} & 
\makebox[0.14\textwidth][c]{$\Sigma_{\mathrm{SFR}}$ [$M_{\odot}/\mathrm{yr}/\mathrm{kpc}^2$]$^e$} \\

\hline
\multicolumn{6}{c}{\textit{Low-mass galaxies ($\log(M_{\ast}/M_{\odot}) < 9$)}} \\
\hline
10636 & $< 8.34$ & $< 8.20$ & \dots & $0.10 \pm 0.01$ & $137.34 \pm 29.42$ \\
11077 & $8.05 \pm 0.06$ & $7.95 \pm 0.06$ & $8.25 \pm 0.11$ & $0.50 \pm 0.03$ & $10.99 \pm 1.48$ \\
11274 & $< 8.49$ & $< 8.34$ & \dots & $0.51 \pm 0.05$ & $6.54 \pm 1.51$ \\
\hline
\multicolumn{6}{c}{\textit{High-mass galaxies ($\log(M_{\ast}/M_{\odot}) > 10$)}} \\
\hline
9523$^{\dagger}$  & $8.48 \pm 0.04$ & $8.58 \pm 0.12$ & $7.31 \pm 0.16$ & $0.84 \pm 0.01$ & $10.61 \pm 1.82$ \\
11833 & $8.41 \pm 0.04$ & $8.54 \pm 0.06$ & \dots & $1.38 \pm 0.04$ & $5.31 \pm 0.83$ \\
12655 & $8.59 \pm 0.05$ & $> 8.49$        & \dots & $2.10 \pm 0.01$ & $3.02 \pm 0.58$ \\
13329 & $8.44 \pm 0.04$ & $8.51 \pm 0.02$ & $7.16 \pm 0.04$ & $2.67 \pm 0.02$ & $1.84 \pm 0.18$ \\
\hline
\multicolumn{6}{l}{\parbox{\textwidth}{\vspace{1ex}
\textbf{Notes.} \\
$^a$ Metallicity ($N2$) was calculated using the $N2$ index following the method of \cite{2013A&A...559A.114M}. \\
$^b$ Metallicity ($O3N2$) was calculated using the $O3N2$ index following the method of \cite{2004MNRAS.348L..59P}. \\
$^c$ $\log q$ was derived using Equation (13) of \cite{Kobulnicky_2004} method, incorporating the $O32$ index and the metallicity derived from the $O3N2$ index. \\
$^d$ The circularized effective radius is defined as $R_{e,circ} = R_{e,major} \times \sqrt{b/a}$, where $R_{e,major}$ is the semi-major axis effective radius and $b/a$ is the axis ratio. \\
$^e$ For high-mass galaxies, the $\Sigma_{\mathrm{SFR}}$ was calculated using star formation rates derived from the $K_s$-band flux excess method, while for the low-mass galaxies, star formation rates were derived following the calibration of \citet{1998ARA&A..36..189K} assuming a \citet{Chabrier_2003} initial mass function (IMF). \\
$^{\dagger}$ Confirmed as an AGN through X-ray cross-matching with the catalog presented by \citet{Cowley_2016}.

}}
\end{tabular*}
\end{table*}

%%%%% Table 4 - (metallicity, log q) - End %%%%%

%%%%%%%%%%%%%%%%%%%%%%%%%%%%%%%%%%%%%%%%%%%%%%%%%%%%%%%%%%%%%%%%%%%%%%%%%%%%%%%%
\section{Discussion} \label{sec:discussion}

\subsection{Validation of the $K_s$-band Flux Excess Method}

The seven galaxies were selected solely on the basis of their $K_s$-band flux excess, without using prior spectroscopic confirmation as a selection criterion. Still, the subsequent spectroscopy confirmed all seven galaxies as HAEs in the redshift range $2.06 < z < 2.49$, supporting the effectiveness of the $K_s$-band flux excess selection method. A comparison between the H$\alpha$ fluxes derived from the $K_s$-band excess and those measured from spectroscopy shows that the two methods yield comparable fluxes when the spectroscopic slit adequately covers the spatial extent of the galaxy. Furthermore, the result demonstrates that the method remains effective even for low-mass galaxies, where the stellar continuum is intrinsically faint. Although the sample size is small, the 100\% spectroscopic confirmation rate supports the use of $K_s$-band flux excess selection as an efficient pre-selection technique for assembling HAE samples, providing a practical pathway for constructing statistically meaningful samples prior to detailed spectroscopic follow-up, especially in wide-area photometric surveys.

%%%%%%%%%%%%%%%%%%%%%%%%%%%%%%%%%%%%%%%%%%%%%%%%%%%%%%%%%%%%%%%%%%%%%%%%%%%%%%%%
\subsection{Physical Properties of the Low-Mass Galaxies}

The low-mass galaxies exhibit distinct spectroscopic characteristics, including weak [N \textsc{ii}] and [O \textsc{ii}] emission lines together with relatively strong [O \textsc{iii}] emission line, which are consistent with highly ionized nebular conditions and low gas-phase metallicities. In contrast, the high-mass galaxies show a higher $N2$ value compared to the low-mass galaxies, suggesting more chemically enriched interstellar media.

Table~\ref{tab:galaxy_properties_metal} summarizes the physical properties of the target galaxies. Metallicities estimated from the $N2$ and $O3N2$ calibrations indicate that the low-mass galaxy ID 11077 yields $12+\log(\text{O/H}) = 7.95$--$8.05$. Also, the upper limits for the metallicity of the two low-mass galaxies remain below the metallicity range measured for the high-mass galaxies of $12+\log(\text{O/H}) = 8.44$--$8.59$. Furthermore, the ionization parameters indicate a systematic difference between the high- and low-mass galaxies: the low-mass galaxy ID 11077 exhibits $\log U = -2.23 \pm 0.14$, which is approximately 1.0~dex higher than that of the high-mass galaxies. This may reflect the combined effects of the compact star-forming structure and lower metallicity in this galaxy. 

Based on the JWST F200W image, the effective radius of the low-mass galaxies is $R_{e,\mathrm{circ}} \sim 0.50$~kpc, smaller than $R_{e,\mathrm{circ}} = 0.84$--$2.67$~kpc for the high-mass galaxies. Such compact morphologies lead to elevated star-formation surface densities. In particular, ID~10636 exhibits an exceptionally high $\Sigma_{\rm SFR}$ of $137.34 \pm 29.42\,M_\odot\,\mathrm{yr^{-1}\,kpc^{-2}}$, exceeding the average values of the high-mass sample ($\sim 5\,M_\odot\,\mathrm{yr^{-1}\,kpc^{-2}}$) by more than an order of magnitude.

However, the available measurements suggest that high $\Sigma_{\rm SFR}$ alone is unlikely to fully explain the observed ionization conditions. This comparison tentatively suggests that metallicity may also play an important role in regulating ionization state, which is consistent with the well-known anti-correlation between the ionization parameter and gas-phase metallicity reported in previous studies \citep[e.g.,][]{1986ApJ...307..431D,2014MNRAS.441.2663P}. Lower-metallicity stellar populations produce harder ionizing photons, which can enhance nebular excitation. In this picture, the combination of chemically young stellar populations and compact star-forming geometry may contribute to sustaining the elevated ionization conditions observed in the low-mass galaxies. Although the present sample is limited, the observed differences between the low- and high-mass galaxies qualitatively follow the trend that lower-mass galaxies tend to have lower metallicities.

%%%%%%%%%%%%%%%%%%%%%%%%%%%%%%%%%%%%%%%%%%%%%%%%%%%%%%%%%%%%%%%%%%%%%%%%

%% Figure7 - F115W-F277W - Start
\begin{figure*}[htbp]
\centering
\includegraphics[width=\textwidth]{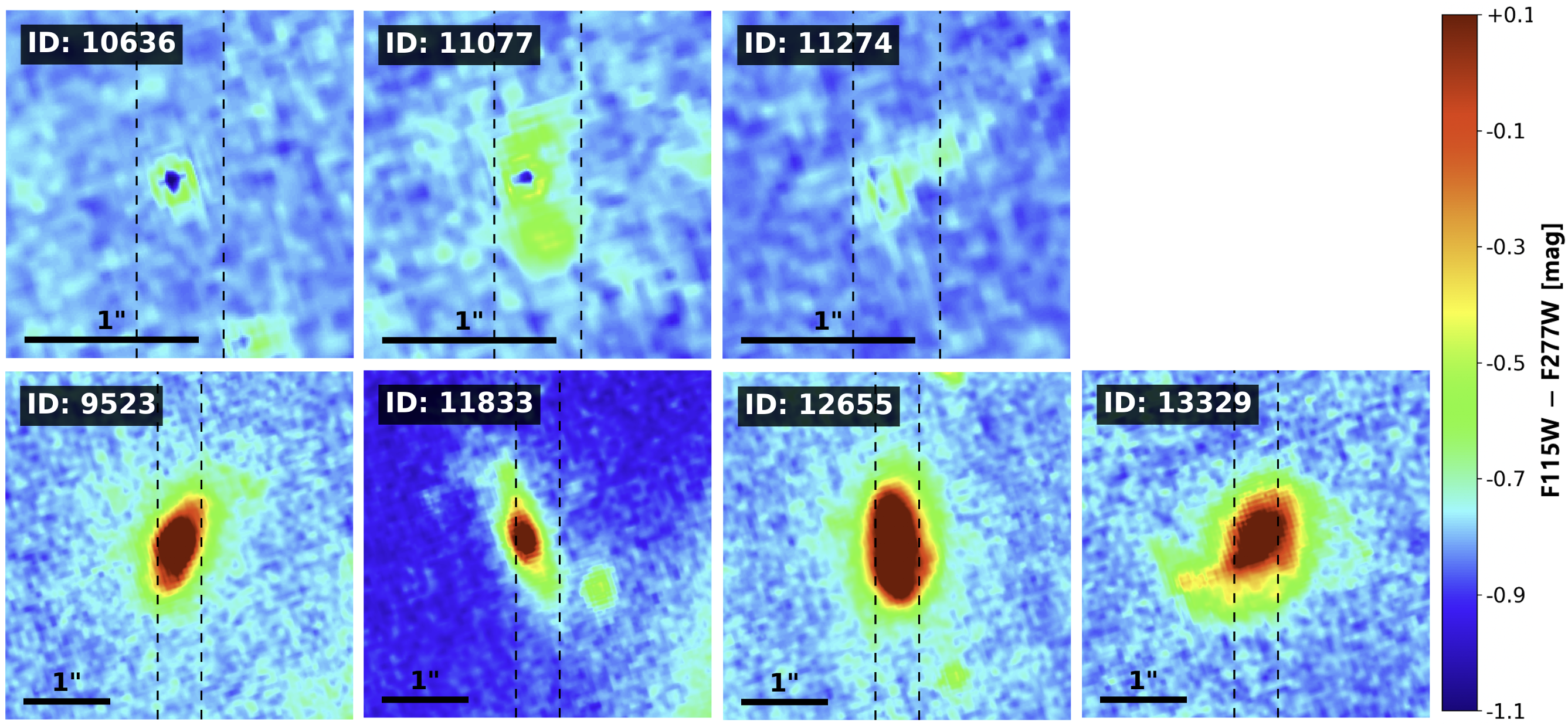} 
\caption{Spatially resolved F115W$-$F277W color maps of the target galaxies. The top row shows the low-mass galaxies (IDs 10636, 11077, 11274), and the bottom row shows the high-mass galaxies (IDs 9523, 11833, 12655, 13329). Each panel includes a $1''$ scale bar and the galaxy ID. The dashed lines indicate the position of the $0.5'' \times 10''$ spectroscopic slit, centered according to the R.A. and Decl. coordinates from the ZFOURGE catalog. The field of view is $2.0''$ for the top row and $4.0''$ for the bottom row. The color scale represents the magnitude difference ($m_{\text{F115W}} - m_{\text{F277W}}$).}
\label{fig:color_maps}
\end{figure*}
%% Figure7 - F115W-F277W - End

%%%%%%%%%%%%%%%%%%%%%%%%%%%%%%%%%%%%%%%%%%%%%%%%%%%%%%%%%%%%%%%%%%%%%%%%%%%

\subsection{Radial Color Gradients}

To probe detailed spatial structure of stellar population, we examined F115W and F277W images of JWST/PRIMER dataset \citep{2021jwst.prop.1837D}, which provides deep coverage of the rest-frame $U$-band and $I$-band emission of $z \sim 2.0$--2.5 galaxies. These data are available through the MAST at \dataset[10.17909/6v45-xr78]{https://doi.org/10.17909/6v45-xr78}. This filter combination traces variations in the optical continuum slope and provides sensitivity to age-dependent spectral features of Balmer break, allowing us to qualitatively assess the spatial distribution of young stellar populations relative to the underlying older component.

To construct reliable spatially resolved F115W--F277W color maps, all images were PSF-matched to a common effective resolution to account for differences between the short- and long-wavelength NIRCam channels. Model PSFs for the F115W and F277W filters were generated using the \texttt{stpsf} package. The higher-resolution F115W image was convolved to the F277W PSF using a kernel generated with the \texttt{pypher} package, with pixel-scale differences between the two channels taken into account to preserve photometric consistency. From the PSF-matched images, we construct spatially resolved F115W-F277W color maps shown in Figure~\ref{fig:color_maps}.

\subsubsection{Positive Gradients in Low-Mass HAEs}

The low-mass galaxies exhibit positive radial color gradients, with bluer central regions compared to their outskirts. Although broadband colors are subject to age--dust degeneracies, low $E(B-V)$ values derived from SED fitting suggest that the blue color at the center may indicate their very young stellar population. In addition, \cite{jin2025highincidencecentralstar} reported that the positive radial color gradients of the low-mass galaxies are less affected by dust reddening. Taken together, we interpret the blue cores as reflecting a genuine concentration of young stellar populations, indicative of centrally concentrated star formation activity.

Such central color gradients are consistent with scenarios in which centrally concentrated gas accumulation—potentially driven by gas inflows or internal disk instabilities—triggers intense, compact star formation episodes \citep{Zolotov_2015} and results in high ionization conditions. This interpretation is further supported by the low gas-phase metallicity and high excitation nebular emission-line properties reported above. This centralized growth behavior may represent an early phase of stellar mass assembly, during which centrally confined star formation drives a rapid increase in stellar surface density.

To further characterize the nature of the blue central regions, we estimated the stellar ages of the bluest pixels in the F115W-F277W color maps for the low-mass galaxies. After correcting for dust attenuation using $E(B-V)$ in Table~\ref{tab:zfourge_properties}, we compared the dereddened colors with a simple stellar population (SSP) model \texttt{bc2003\_lr\_BaSeL\_m42\_chab\_ssp.ised} with a metallicity of $Z=0.2\,Z_\odot$ ($12+\log(\mathrm{O/H}) \approx 8.0$) from \cite{Bruzual_2003}, which is consistent with that of the low-mass galaxies.

The derived ages are $7.2$~Myr for ID~10636, $8.7$~Myr for ID~11077, and $10.5$~Myr for ID~11274, supporting the interpretation that the blue cores are dominated by recent, intense star formation episodes occurring on short timescales.

Within this picture, the low-mass galaxies can be interpreted as systems observed during episodes of enhanced central star formation, which is consistent with theoretical expectations for the low-mass galaxies where shallow gravitational potentials and strong stellar feedback lead to bursty and time-variable star-formation histories \citep[e.g.,][]{Weisz_2011, Hopkins_2014}.

We cannot entirely exclude the possibility that residual systematics associated with the PSF matching between the F115W and F277W images may affect the measured color gradients, particularly for the low-mass galaxies. Because these galaxies have relatively small angular sizes, residual PSF mismatch could smooth or dilute intrinsic color gradients. However, the same PSF-matching procedure was uniformly applied to all galaxies, while centrally red color gradients are consistently observed only in the high-mass galaxies. Therefore, although residual PSF mismatch may slightly affect the absolute gradient amplitude in the low-mass galaxies, it is unlikely to explain the systematic difference in radial color gradients between the low- and high-mass galaxies.

\subsubsection{Negative Gradients in High-Mass HAEs}

The high-mass galaxies in Figure~\ref{fig:color_maps} display negative radial color gradients, characterized by redder cores and progressively bluer outskirts. This has often been referred to as inside-out growth, in which the central regions are dominated by more evolved stellar populations and/or enhanced dust attenuation, while ongoing star formation is extended over larger spatial scales in the outskirts \citep{Tacchella_2015, Nelson_2016}.

In addition to their negative color gradients, the high-mass galaxies are suggested to have extended H$\alpha$ distributions, which is consistent with previous studies showing that massive star-forming galaxies commonly host extended stellar structures and disk-wide star formation (e.g., \citealt{Nelson_2016}).

Together with the higher metallicity observed in the high-mass galaxies, our results reinforce the picture in which massive star-forming galaxies at cosmic noon have already developed mature central components, while continuing to build stellar mass through extended star formation in their outer regions.

\subsection{Ly$\alpha$ Emission and Escape Fraction}

All of the low-mass galaxies have been spectroscopically confirmed as Ly$\alpha$ emitters (LAEs) in \cite{2020ApJ...903...24M} and \cite{2014ApJ...796...64C}, whereas no information on Ly$\alpha$ emission was available for the high-mass galaxies. Previous studies have shown that $z \sim 2$ LAEs in narrow-band surveys span a wide range of stellar masses and are not systematically biased toward the low-mass regime \citep{2011A&A...529A...9N, 2011ApJ...733..114G, Vargas_2014, 2018PASJ...70....4K}. They typically exhibit low dust attenuation, $E(B\!-\!V) \sim 0.0$--0.3 \citep{2011ApJ...733..114G, Vargas_2014}, which is comparable to the low-mass galaxies.

We estimated the Ly$\alpha$ escape fraction, which is defined as the ratio of the observed Ly$\alpha$ flux to the intrinsic Ly$\alpha$ flux produced within the galaxy. The escape fraction is expressed as \citep{2015PASA...32...27H}:
\begin{equation}
f_{\mathrm{esc}}(\mathrm{Ly}\alpha)
= \frac{F_{\mathrm{obs}}(\mathrm{Ly}\alpha)}
{F_{\mathrm{int}}(\mathrm{Ly}\alpha)}
= \frac{F_{\mathrm{obs}}(\mathrm{Ly}\alpha)}
{8.7  F_{\mathrm{corr}}(\mathrm{H}\alpha)}
\end{equation}
where $F_{\mathrm{obs}}(\mathrm{Ly}\alpha)$ and $F_{\mathrm{int}}(\mathrm{Ly}\alpha)$ are the observed and intrinsic $\mathrm{Ly}\alpha$ fluxes, respectively, and $F_{\mathrm{corr}}(\mathrm{H}\alpha)$ is the $\mathrm{H}\alpha$ flux corrected for dust extinction. The factor 8.7 represents the intrinsic $\mathrm{Ly}\alpha/\mathrm{H}\alpha$ ratio under typical nebular conditions ($T \sim 10^4$~K, $n_e \sim 10^2$~cm$^{-3}$). We obtain $f_{\mathrm{esc}}(\mathrm{Ly}\alpha) = 0.977^{+0.313}_{-0.217}$, $1.420^{+0.340}_{-0.160}$, and $0.443^{+0.051}_{-0.051}$ for IDs 10636, 11077, and 11274, respectively. 

The inferred escape fraction of ID 11077 exceeds unity. This value should be interpreted with caution, as it may arise from uncertainties in the Ly$\alpha$ and H$\alpha$ flux measurements, the dust correction, aperture effects, or deviations from the assumed intrinsic Ly$\alpha$/H$\alpha$ ratio. Overall, the low-mass galaxies show high inferred Ly$\alpha$ escape fractions, with $f_{\mathrm{esc}}(\mathrm{Ly}\alpha)=0.44$--0.98 for IDs 10636 and 11274. These inferred escape fractions are higher than those found for narrow-band-selected LAE populations, for which $f_{\mathrm{esc}}(\mathrm{Ly}\alpha) \sim 12\text{--}30\%$ have been reported \citep{2012ApJ...745...12N}. This may indicate that the low-mass galaxies have physical conditions favorable for Ly$\alpha$ escape, potentially associated with their low dust attenuation, high ionization states, and porous or clumpy interstellar medium geometries.

\section{Summary} 
\label{sec:summary}

In this study, we presented a spectroscopic analysis and spatial distribution of seven HAE candidates at $z \sim 2$ in the ZFOURGE--COSMOS field. By combining deep Subaru/SWIMS near-infrared spectroscopy with high-resolution JWST/NIRCam imaging, we characterized the physical properties of target galaxies across two distinct mass regimes. Our primary findings are summarized as follows:

\begin{enumerate}
    \item Efficiency of the $K_s$-band Flux Excess Method: We spectroscopically confirmed H$\alpha$ emission line in all seven candidates at $z=2.06$--2.49, achieving a 100\% selection efficiency within our observed sample. This result validates the $K_s$-band flux excess method as a highly reliable tool for identifying star-forming galaxies at cosmic noon, particularly reaching down to the low-mass regime that is often underrepresented in large spectroscopic surveys.

    \item Compact vs. Extended Star-Formation Distribution: We identified a mass-dependent discrepancy in H$\alpha$ flux measurements. The low-mass galaxies show consistent H$\alpha$ fluxes between photometry and spectroscopy, indicating that the primary star-forming regions lie within the slit aperture. High-mass galaxies exhibit higher photometric fluxes due to slit-loss caused by extended star-formation distributions. This morphological distinction is further confirmed by JWST/NIRCam F200W imaging, which reveals compact structures in low-mass galaxies and extended, complex structures in high-mass systems.

    \item Mass-Dependent Nebular Properties: The sample exhibits a clear mass-dependent difference in nebular excitation and chemical enrichment. The low-mass galaxies are identified as EELGs, exhibiting [O\,\textsc{iii}]~$\lambda5007$ fluxes stronger than H$\alpha$, high $R3$ and $O32$ ratios, and weak [N\,\textsc{ii}]~$\lambda6584$ emission line, indicating high ionization parameters and low gas-phase metallicities. In contrast, the high-mass galaxies exhibit H$\alpha$ fluxes stronger than [O\,\textsc{iii}]~$\lambda5007$, lower $O32$ ratios, and stronger [N\,\textsc{ii}]~$\lambda6584$ emission line, consistent with lower ionization parameters and higher gas-phase metallicities.

    \item Reionization-Epoch Analogs at Cosmic Noon: The low-mass galaxies at $z \sim 2$ occupy similar regions in the $O32$--$R23$ and $R3$--$R2$ diagnostic diagrams to those of local Green Pea and Blueberry galaxies, as well as high-redshift star-forming galaxies at $z \gtrsim 6$. They also exhibit strong [O\,\textsc{iii}]~$\lambda5007$ emission line, high sSFRs, low dust reddening, and compact morphologies. These similarities indicate that low-mass high-sSFR galaxies at $z\sim2$ share many of the physical characteristics of reionization-epoch galaxies.

    \item Contrasting Color Gradients: Spatially resolved F115W-F277W color maps reveal distinct radial gradients that vary with stellar mass. The low-mass galaxies show blue cores with positive gradients, consistent with centrally concentrated star formation. Their central regions are indicative of very young stellar populations, and they exhibit globally low dust attenuation. In contrast, high-mass targets show red cores with negative gradients, consistent with the inside-out growth scenarios.

\end{enumerate}

Overall, our results highlight that galaxies at cosmic noon assemble their stellar mass through mass-dependent evolutionary pathways, rather than following a single evolutionary track. Although our findings provide evidence for these mass-dependent growth modes, they are limited by a small sample size and a lack of kinematic data. Future IFU observations will be essential for mapping metallicity gradients and gas kinematics, providing strong constraints on the feedback-regulated starburst and gas inflow scenarios proposed in this work.
%%%%%%%%%%%%%%%%%%%%%%%%%%%%%%%%%%%%%%%%%%%%%%%%%%%%%%%%%%%%%%%%%%%%%%%%%%%%%%%%

\section{Acknowledgments}
This research is based on data collected at the Subaru Telescope, which is operated by the National Astronomical Observatory of Japan. We are honored and grateful for the opportunity of observing the Universe from Maunakea, which has the cultural, historical, and natural significance in Hawaii. This work is also based in part on observations taken by the 3D-HST Treasury Program (GO 12177 and 12328) with the NASA/ESA HST, which is operated by the Association of Universities for Research in Astronomy, Inc., under NASA contract NAS5-26555. Data analysis was in part carried out on the Multi-wavelength Data Analysis System operated by the Astronomy Data Center (ADC), National Astronomical Observatory of Japan. This work was supported by JSPS Core-to-Core Program (grant number: JPJSCCA20210003) and JSPS KAKENHI Grant Numbers JP15H02062, JP20H00171, JP22K21349, JP24H00002, and JP25H00663.

%% For this sample we use BibTeX plus aasjournalv7.bst to generate the

\bibliography{sample701}{}

@article{Madau_2014,
   title={Cosmic Star-Formation History},
   volume={52},
   ISSN={1545-4282},
   url={http://dx.doi.org/10.1146/annurev-astro-081811-125615},
   DOI={10.1146/annurev-astro-081811-125615},
   number={1},
   journal={Annual Review of Astronomy and Astrophysics},
   publisher={Annual Reviews},
   author={Madau, Piero and Dickinson, Mark},
   year={2014},
   month=aug, pages={415–486} }

@article{Kere__2005,
   title={How do galaxies get their gas?},
   volume={363},
   ISSN={1365-2966},
   url={http://dx.doi.org/10.1111/j.1365-2966.2005.09451.x},
   DOI={10.1111/j.1365-2966.2005.09451.x},
   number={1},
   journal={Monthly Notices of the Royal Astronomical Society},
   publisher={Oxford University Press (OUP)},
   author={Kere , D. and Katz, N. and Weinberg, D. H. and Dave, R.},
   year={2005},
   month=oct, pages={2–28} }

@article{Dekel_2009,
   title={Cold streams in early massive hot haloes as the main mode of galaxy formation},
   volume={457},
   ISSN={1476-4687},
   url={http://dx.doi.org/10.1038/nature07648},
   DOI={10.1038/nature07648},
   number={7228},
   journal={Nature},
   publisher={Springer Science and Business Media LLC},
   author={Dekel, A. and Birnboim, Y. and Engel, G. and Freundlich, J. and Goerdt, T. and Mumcuoglu, M. and Neistein, E. and Pichon, C. and Teyssier, R. and Zinger, E.},
   year={2009},
   month=jan, pages={451–454} }

@article{Somerville_2015,
   title={Star formation in semi-analytic galaxy formation models with multiphase gas},
   volume={453},
   ISSN={1365-2966},
   url={http://dx.doi.org/10.1093/mnras/stv1877},
   DOI={10.1093/mnras/stv1877},
   number={4},
   journal={Monthly Notices of the Royal Astronomical Society},
   publisher={Oxford University Press (OUP)},
   author={Somerville, Rachel S. and Popping, Gergö and Trager, Scott C.},
   year={2015},
   month=sep, pages={4338–4368} }

@article{Daddi_2007,
   title={Multiwavelength Study of Massive Galaxies atz∼2. I. Star Formation and Galaxy Growth},
   volume={670},
   ISSN={1538-4357},
   url={http://dx.doi.org/10.1086/521818},
   DOI={10.1086/521818},
   number={1},
   journal={The Astrophysical Journal},
   publisher={American Astronomical Society},
   author={Daddi, E. and Dickinson, M. and Morrison, G. and Chary, R. and Cimatti, A. and Elbaz, D. and Frayer, D. and Renzini, A. and Pope, A. and Alexander, D. M. and Bauer, F. E. and Giavalisco, M. and Huynh, M. and Kurk, J. and Mignoli, M.},
   year={2007},
   month=nov, pages={156–172} }

@article{Rodighiero_2011,
   title={THE LESSER ROLE OF STARBURSTS IN STAR FORMATION AT
                    z
                    = 2},
   volume={739},
   ISSN={2041-8213},
   url={http://dx.doi.org/10.1088/2041-8205/739/2/L40},
   DOI={10.1088/2041-8205/739/2/l40},
   number={2},
   journal={The Astrophysical Journal},
   publisher={American Astronomical Society},
   author={Rodighiero, G. and Daddi, E. and Baronchelli, I. and Cimatti, A. and Renzini, A. and Aussel, H. and Popesso, P. and Lutz, D. and Andreani, P. and Berta, S. and Cava, A. and Elbaz, D. and Feltre, A. and Fontana, A. and Förster Schreiber, N. M. and Franceschini, A. and Genzel, R. and Grazian, A. and Gruppioni, C. and Ilbert, O. and Le Floch, E. and Magdis, G. and Magliocchetti, M. and Magnelli, B. and Maiolino, R. and McCracken, H. and Nordon, R. and Poglitsch, A. and Santini, P. and Pozzi, F. and Riguccini, L. and Tacconi, L. J. and Wuyts, S. and Zamorani, G.},
   year={2011},
   month=sep, pages={L40} }

@article{Speagle_2014,
   title={A HIGHLY CONSISTENT FRAMEWORK FOR THE EVOLUTION OF THE STAR-FORMING “MAIN SEQUENCE” FROM
                    z
                    ∼ 0-6},
   volume={214},
   ISSN={1538-4365},
   url={http://dx.doi.org/10.1088/0067-0049/214/2/15},
   DOI={10.1088/0067-0049/214/2/15},
   number={2},
   journal={The Astrophysical Journal Supplement Series},
   publisher={American Astronomical Society},
   author={Speagle, J. S. and Steinhardt, C. L. and Capak, P. L. and Silverman, J. D.},
   year={2014},
   month=sep, pages={15} }

@article{Chen_2024,
   title={Multiple Emission Lines of Hα Emitters at z ∼ 2.3 from the Broad- and Medium-band Photometry in the ZFOURGE Survey},
   volume={964},
   ISSN={1538-4357},
   url={http://dx.doi.org/10.3847/1538-4357/ad20cc},
   DOI={10.3847/1538-4357/ad20cc},
   number={1},
   journal={The Astrophysical Journal},
   publisher={American Astronomical Society},
   author={Chen, Nuo and Motohara, Kentaro and Spitler, Lee and Nakajima, Kimihiko and Momose, Rieko and Kodama, Tadayuki and Konishi, Masahiro and Takahashi, Hidenori and Kushibiki, Kosuke and Kono, Yukihiro and Terao, Yasunori},
   year={2024},
   month=mar, pages={5} }

@article{Chabrier_2003,
   title={Galactic Stellar and Substellar Initial Mass Function},
   volume={115},
   ISSN={1538-3873},
   url={http://dx.doi.org/10.1086/376392},
   DOI={10.1086/376392},
   number={809},
   journal={Publications of the Astronomical Society of the Pacific},
   publisher={IOP Publishing},
   author={Chabrier, Gilles},
   year={2003},
   month=jul, pages={763–795} }

@article{Straatman_2016,
   title={THE FOURSTAR GALAXY EVOLUTION SURVEY (ZFOURGE): ULTRAVIOLET TO FAR-INFRARED CATALOGS, MEDIUM-BANDWIDTH PHOTOMETRIC REDSHIFTS WITH IMPROVED ACCURACY, STELLAR MASSES, AND CONFIRMATION OF QUIESCENT GALAXIES TO z ∼ 3.5*},
   volume={830},
   ISSN={1538-4357},
   url={http://dx.doi.org/10.3847/0004-637X/830/1/51},
   DOI={10.3847/0004-637x/830/1/51},
   number={1},
   journal={The Astrophysical Journal},
   publisher={American Astronomical Society},
   author={Straatman, Caroline M. S. and Spitler, Lee R. and Quadri, Ryan F. and Labbé, Ivo and Glazebrook, Karl and Persson, S. Eric and Papovich, Casey and Tran, Kim-Vy H. and Brammer, Gabriel B. and Cowley, Michael and Tomczak, Adam and Nanayakkara, Themiya and Alcorn, Leo and Allen, Rebecca and Broussard, Adam and van Dokkum, Pieter and Forrest, Ben and van Houdt, Josha and Kacprzak, Glenn G. and Kawinwanichakij, Lalitwadee and Kelson, Daniel D. and Lee, Janice and McCarthy, Patrick J. and Mehrtens, Nicola and Monson, Andrew and Murphy, David and Rees, Glen and Tilvi, Vithal and Whitaker, Katherine E.},
   year={2016},
   month=oct, pages={51} }

@software{2016ascl.soft06014N,
       author = {{Newville}, Matthew and {Stensitzki}, Till and {Allen}, Daniel B. and {Rawlik}, Michal and {Ingargiola}, Antonino and {Nelson}, Andrew},
        title = "{Lmfit: Non-Linear Least-Square Minimization and Curve-Fitting for Python}",
 howpublished = {Astrophysics Source Code Library, record ascl:1606.014},
         year = 2016,
        month = jun,
          eid = {ascl:1606.014},
archivePrefix = {ascl},
       eprint = {1606.014},
       adsurl = {https://ui.adsabs.harvard.edu/abs/2016ascl.soft06014N}
}

@ARTICLE{2025ApJ...982...27Z,
       author = {{Zhu}, Chenghao and {Harikane}, Yuichi and {Ouchi}, Masami and {Ono}, Yoshiaki and {Onodera}, Masato and {Tang}, Shenli and {Isobe}, Yuki and {Matsuoka}, Yoshiki and {Kawaguchi}, Toshihiro and {Umeda}, Hiroya and {Nakajima}, Kimihiko and {Liang}, Yongming and {Xu}, Yi and {Zhang}, Yechi and {Sun}, Dongsheng and {Shimasaku}, Kazuhiro and {Greene}, Jenny and {Iwasawa}, Kazushi and {Kohno}, Kotaro and {Nagao}, Tohru and {Schulze}, Andreas and {Shibuya}, Takatoshi and {Hilmi}, Miftahul and {Schramm}, Malte},
        title = "{The Physical Origin of Extreme Emission Line Galaxies at High Redshifts: Strong [O III] Emission Lines Produced by Obscured Active Galactic Nuclei}",
      journal = {\apj},
         year = 2025,
        month = mar,
       volume = {982},
       number = {1},
          eid = {27},
        pages = {27},
          doi = {10.3847/1538-4357/adaf1b},
archivePrefix = {arXiv},
       eprint = {2410.12198},
 primaryClass = {astro-ph.GA},
       adsurl = {https://ui.adsabs.harvard.edu/abs/2025ApJ...982...27Z}
}

@ARTICLE{2000ApJ...533..682C,
       author = {{Calzetti}, Daniela and {Armus}, Lee and {Bohlin}, Ralph C. and {Kinney}, Anne L. and {Koornneef}, Jan and {Storchi-Bergmann}, Thaisa},
        title = "{The Dust Content and Opacity of Actively Star-forming Galaxies}",
      journal = {\apj},
         year = 2000,
        month = apr,
       volume = {533},
       number = {2},
        pages = {682-695},
          doi = {10.1086/308692},
archivePrefix = {arXiv},
       eprint = {astro-ph/9911459},
 primaryClass = {astro-ph},
       adsurl = {https://ui.adsabs.harvard.edu/abs/2000ApJ...533..682C}
}

@ARTICLE{2020ApJ...902..123R,
       author = {{Reddy}, Naveen A. and {Shapley}, Alice E. and {Kriek}, Mariska and {Steidel}, Charles C. and {Shivaei}, Irene and {Sanders}, Ryan L. and {Mobasher}, Bahram and {Coil}, Alison L. and {Siana}, Brian and {Freeman}, William R. and {Azadi}, Mojegan and {Fetherolf}, Tara and {Leung}, Gene and {Price}, Sedona H. and {Zick}, Tom},
        title = "{The MOSDEF Survey: The First Direct Measurements of the Nebular Dust Attenuation Curve at High Redshift}",
      journal = {\apj},
         year = 2020,
        month = oct,
       volume = {902},
       number = {2},
          eid = {123},
        pages = {123},
          doi = {10.3847/1538-4357/abb674},
archivePrefix = {arXiv},
       eprint = {2009.10085},
 primaryClass = {astro-ph.GA},
       adsurl = {https://ui.adsabs.harvard.edu/abs/2020ApJ...902..123R}
}

@ARTICLE{2013ApJ...777L...8K,
       author = {{Kashino}, D. and {Silverman}, J.~D. and {Rodighiero}, G. and {Renzini}, A. and {Arimoto}, N. and {Daddi}, E. and {Lilly}, S.~J. and {Sanders}, D.~B. and {Kartaltepe}, J. and {Zahid}, H.~J. and {Nagao}, T. and {Sugiyama}, N. and {Capak}, P. and {Carollo}, C.~M. and {Chu}, J. and {Hasinger}, G. and {Ilbert}, O. and {Kajisawa}, M. and {Kewley}, L.~J. and {Koekemoer}, A.~M. and {Kova{\v{c}}}, K. and {Le F{\`e}vre}, O. and {Masters}, D. and {McCracken}, H.~J. and {Onodera}, M. and {Scoville}, N. and {Strazzullo}, V. and {Symeonidis}, M. and {Taniguchi}, Y.},
        title = "{The FMOS-COSMOS Survey of Star-forming Galaxies at z \raisebox{-0.5ex}\textasciitilde 1.6. I. H{\ensuremath{\alpha}}-based Star Formation Rates and Dust Extinction}",
      journal = {\apjl},
         year = 2013,
        month = nov,
       volume = {777},
       number = {1},
          eid = {L8},
        pages = {L8},
          doi = {10.1088/2041-8205/777/1/L8},
archivePrefix = {arXiv},
       eprint = {1309.4774},
 primaryClass = {astro-ph.CO},
       adsurl = {https://ui.adsabs.harvard.edu/abs/2013ApJ...777L...8K}
}

@ARTICLE{1998ARA&A..36..189K,
       author = {{Kennicutt}, Jr., Robert C.},
        title = "{Star Formation in Galaxies Along the Hubble Sequence}",
      journal = {\araa},
         year = 1998,
        month = jan,
       volume = {36},
        pages = {189-232},
          doi = {10.1146/annurev.astro.36.1.189},
archivePrefix = {arXiv},
       eprint = {astro-ph/9807187},
 primaryClass = {astro-ph},
       adsurl = {https://ui.adsabs.harvard.edu/abs/1998ARA&A..36..189K}
}

@ARTICLE{2012ApJS..200...13B,
       author = {{Brammer}, Gabriel B. and {van Dokkum}, Pieter G. and {Franx}, Marijn and {Fumagalli}, Mattia and {Patel}, Shannon and {Rix}, Hans-Walter and {Skelton}, Rosalind E. and {Kriek}, Mariska and {Nelson}, Erica and {Schmidt}, Kasper B. and {Bezanson}, Rachel and {da Cunha}, Elisabete and {Erb}, Dawn K. and {Fan}, Xiaohui and {F{\"o}rster Schreiber}, Natascha and {Illingworth}, Garth D. and {Labb{\'e}}, Ivo and {Leja}, Joel and {Lundgren}, Britt and {Magee}, Dan and {Marchesini}, Danilo and {McCarthy}, Patrick and {Momcheva}, Ivelina and {Muzzin}, Adam and {Quadri}, Ryan and {Steidel}, Charles C. and {Tal}, Tomer and {Wake}, David and {Whitaker}, Katherine E. and {Williams}, Anna},
        title = "{3D-HST: A Wide-field Grism Spectroscopic Survey with the Hubble Space Telescope}",
      journal = {\apjs},
         year = 2012,
        month = jun,
       volume = {200},
       number = {2},
          eid = {13},
        pages = {13},
          doi = {10.1088/0067-0049/200/2/13},
archivePrefix = {arXiv},
       eprint = {1204.2829},
 primaryClass = {astro-ph.CO},
       adsurl = {https://ui.adsabs.harvard.edu/abs/2012ApJS..200...13B}
}

@ARTICLE{2016ApJS..225...27M,
       author = {{Momcheva}, Ivelina G. and {Brammer}, Gabriel B. and {van Dokkum}, Pieter G. and {Skelton}, Rosalind E. and {Whitaker}, Katherine E. and {Nelson}, Erica J. and {Fumagalli}, Mattia and {Maseda}, Michael V. and {Leja}, Joel and {Franx}, Marijn and {Rix}, Hans-Walter and {Bezanson}, Rachel and {Da Cunha}, Elisabete and {Dickey}, Claire and {F{\"o}rster Schreiber}, Natascha M. and {Illingworth}, Garth and {Kriek}, Mariska and {Labb{\'e}}, Ivo and {Ulf Lange}, Johannes and {Lundgren}, Britt F. and {Magee}, Daniel and {Marchesini}, Danilo and {Oesch}, Pascal and {Pacifici}, Camilla and {Patel}, Shannon G. and {Price}, Sedona and {Tal}, Tomer and {Wake}, David A. and {van der Wel}, Arjen and {Wuyts}, Stijn},
        title = "{The 3D-HST Survey: Hubble Space Telescope WFC3/G141 Grism Spectra, Redshifts, and Emission Line Measurements for \raisebox{-0.5ex}\textasciitilde 100,000 Galaxies}",
      journal = {\apjs},
         year = 2016,
        month = aug,
       volume = {225},
       number = {2},
          eid = {27},
        pages = {27},
          doi = {10.3847/0067-0049/225/2/27},
archivePrefix = {arXiv},
       eprint = {1510.02106},
 primaryClass = {astro-ph.GA},
       adsurl = {https://ui.adsabs.harvard.edu/abs/2016ApJS..225...27M}
}

@ARTICLE{2016ApJ...828...21N,
       author = {{Nanayakkara}, Themiya and {Glazebrook}, Karl and {Kacprzak}, Glenn G. and {Yuan}, Tiantian and {Tran}, Kim-Vy and {Spitler}, Lee and {Kewley}, Lisa and {Straatman}, Caroline and {Cowley}, Michael and {Fisher}, David and {Labbe}, Ivo and {Tomczak}, Adam and {Allen}, Rebecca and {Alcorn}, Leo},
        title = "{ZFIRE: A KECK/MOSFIRE Spectroscopic Survey of Galaxies in Rich Environments at z {\ensuremath{\sim}} 2}",
      journal = {\apj},
         year = 2016,
        month = sep,
       volume = {828},
       number = {1},
          eid = {21},
        pages = {21},
          doi = {10.3847/0004-637X/828/1/21},
archivePrefix = {arXiv},
       eprint = {1607.00013},
 primaryClass = {astro-ph.GA},
       adsurl = {https://ui.adsabs.harvard.edu/abs/2016ApJ...828...21N}
}

@ARTICLE{2020ApJ...903...24M,
       author = {{Mukae}, Shiro and {Ouchi}, Masami and {Hill}, Gary J. and {Gebhardt}, Karl and {Cooper}, Erin Mentuch and {Jeong}, Donghui and {Saito}, Shun and {Fabricius}, Maximilian and {Gawiser}, Eric and {Ciardullo}, Robin and {Farrow}, Daniel and {Davis}, Dustin and {Zeimann}, Greg and {Finkelstein}, Steven L. and {Gronwall}, Caryl and {Liu}, Chenxu and {Zhang}, Yechi and {Byrohl}, Chris and {Ono}, Yoshiaki and {Schneider}, Donald P. and {Jarvis}, Matt J. and {Casey}, Caitlin M. and {Mawatari}, Ken},
        title = "{Cosmological 3D H I Gas Map with HETDEX Ly{\ensuremath{\alpha}} Emitters and eBOSS QSOs at z = 2: IGM-Galaxy/QSO Connection and a {\ensuremath{\sim}}40 Mpc Scale Giant H II Bubble Candidate}",
      journal = {\apj},
         year = 2020,
        month = nov,
       volume = {903},
       number = {1},
          eid = {24},
        pages = {24},
          doi = {10.3847/1538-4357/abb81b},
archivePrefix = {arXiv},
       eprint = {2009.07285},
 primaryClass = {astro-ph.GA},
       adsurl = {https://ui.adsabs.harvard.edu/abs/2020ApJ...903...24M}
}

@ARTICLE{2014ApJ...796...64C,
       author = {{Ciardullo}, Robin and {Zeimann}, Gregory R. and {Gronwall}, Caryl and {Gebhardt}, Henry and {Schneider}, Donald P. and {Hagen}, Alex and {Malz}, A.~I. and {Blanc}, Guillermo A. and {Hill}, Gary J. and {Drory}, Niv and {Gawiser}, Eric},
        title = "{Hubble Space Telescope Emission Line Galaxies at z \raisebox{-0.5ex}\textasciitilde 2: The Ly{\ensuremath{\alpha}} Escape Fraction}",
      journal = {\apj},
         year = 2014,
        month = nov,
       volume = {796},
       number = {1},
          eid = {64},
        pages = {64},
          doi = {10.1088/0004-637X/796/1/64},
archivePrefix = {arXiv},
       eprint = {1409.8304},
 primaryClass = {astro-ph.GA},
       adsurl = {https://ui.adsabs.harvard.edu/abs/2014ApJ...796...64C}
}

@article{Cowley_2016,
   title={ZFOURGE catalogue of AGN candidates: an enhancement of 160-μm-derived star formation rates in active galaxies toz = 3.2},
   volume={457},
   ISSN={1365-2966},
   url={http://dx.doi.org/10.1093/mnras/stv2992},
   DOI={10.1093/mnras/stv2992},
   number={1},
   journal={Monthly Notices of the Royal Astronomical Society},
   publisher={Oxford University Press (OUP)},
   author={Cowley, Michael J. and Spitler, Lee R. and Tran, Kim-Vy H. and Rees, Glen A. and Labbé, Ivo and Allen, Rebecca J. and Brammer, Gabriel B. and Glazebrook, Karl and Hopkins, Andrew M. and Juneau, Stéphanie and Kacprzak, Glenn G. and Mullaney, James R. and Nanayakkara, Themiya and Papovich, Casey and Quadri, Ryan F. and Straatman, Caroline M. S. and Tomczak, Adam R. and van Dokkum, Pieter G.},
   year={2016},
   month=jan, pages={629–641} }

@article{del_Moral_Castro_2024,
   title={Extreme-emission-line galaxies in the MUSE Hubble Ultra Deep Field Survey},
   volume={688},
   ISSN={1432-0746},
   url={http://dx.doi.org/10.1051/0004-6361/202348185},
   DOI={10.1051/0004-6361/202348185},
   journal={Astronomy \& Astrophysics},
   publisher={EDP Sciences},
   author={del Moral-Castro, I. and Vílchez, J. M. and Iglesias-Páramo, J. and Arroyo-Polonio, A.},
   year={2024},
   month=jul, pages={A28} }

@ARTICLE{2013ApJ...774L..10K,
       author = {{Kewley}, Lisa J. and {Maier}, Christian and {Yabe}, Kiyoto and {Ohta}, Kouji and {Akiyama}, Masayuki and {Dopita}, Michael A. and {Yuan}, Tiantian},
        title = "{The Cosmic BPT Diagram: Confronting Theory with Observations}",
      journal = {\apjl},
         year = 2013,
        month = sep,
       volume = {774},
       number = {1},
          eid = {L10},
        pages = {L10},
          doi = {10.1088/2041-8205/774/1/L10},
archivePrefix = {arXiv},
       eprint = {1307.0514},
 primaryClass = {astro-ph.CO},
       adsurl = {https://ui.adsabs.harvard.edu/abs/2013ApJ...774L..10K}
}

@ARTICLE{2003MNRAS.346.1055K,
       author = {{Kauffmann}, Guinevere and {Heckman}, Timothy M. and {Tremonti}, Christy and {Brinchmann}, Jarle and {Charlot}, St{\'e}phane and {White}, Simon D.~M. and {Ridgway}, Susan E. and {Brinkmann}, Jon and {Fukugita}, Masataka and {Hall}, Patrick B. and {Ivezi{\'c}}, {\v{Z}}eljko and {Richards}, Gordon T. and {Schneider}, Donald P.},
        title = "{The host galaxies of active galactic nuclei}",
      journal = {\mnras},
         year = 2003,
        month = dec,
       volume = {346},
       number = {4},
        pages = {1055-1077},
          doi = {10.1111/j.1365-2966.2003.07154.x},
archivePrefix = {arXiv},
       eprint = {astro-ph/0304239},
 primaryClass = {astro-ph},
       adsurl = {https://ui.adsabs.harvard.edu/abs/2003MNRAS.346.1055K}
}

@ARTICLE{2004ApJ...613..898T,
       author = {{Tremonti}, Christy A. and {Heckman}, Timothy M. and {Kauffmann}, Guinevere and {Brinchmann}, Jarle and {Charlot}, St{\'e}phane and {White}, Simon D.~M. and {Seibert}, Mark and {Peng}, Eric W. and {Schlegel}, David J. and {Uomoto}, Alan and {Fukugita}, Masataka and {Brinkmann}, Jon},
        title = "{The Origin of the Mass-Metallicity Relation: Insights from 53,000 Star-forming Galaxies in the Sloan Digital Sky Survey}",
      journal = {\apj},
         year = 2004,
        month = oct,
       volume = {613},
       number = {2},
        pages = {898-913},
          doi = {10.1086/423264},
archivePrefix = {arXiv},
       eprint = {astro-ph/0405537},
 primaryClass = {astro-ph},
       adsurl = {https://ui.adsabs.harvard.edu/abs/2004ApJ...613..898T}
}

@ARTICLE{2006ApJ...644..813E,
       author = {{Erb}, Dawn K. and {Shapley}, Alice E. and {Pettini}, Max and {Steidel}, Charles C. and {Reddy}, Naveen A. and {Adelberger}, Kurt L.},
        title = "{The Mass-Metallicity Relation at z>\raisebox{-0.5ex}\textasciitilde2}",
      journal = {\apj},
         year = 2006,
        month = jun,
       volume = {644},
       number = {2},
        pages = {813-828},
          doi = {10.1086/503623},
archivePrefix = {arXiv},
       eprint = {astro-ph/0602473},
 primaryClass = {astro-ph},
       adsurl = {https://ui.adsabs.harvard.edu/abs/2006ApJ...644..813E}
}

@article{Kobulnicky_2004,
   title={Metallicities of 0.3 &lt; z &lt; 1.0 Galaxies in the GOODS‐North Field},
   volume={617},
   ISSN={1538-4357},
   url={http://dx.doi.org/10.1086/425299},
   DOI={10.1086/425299},
   number={1},
   journal={The Astrophysical Journal},
   publisher={American Astronomical Society},
   author={Kobulnicky, Henry A. and Kewley, Lisa J.},
   year={2004},
   month=dec, pages={240–261} }

@article{Hopkins_2014,
   title={Galaxies on FIRE (Feedback In Realistic Environments): stellar feedback explains cosmologically inefficient star formation},
   volume={445},
   ISSN={1365-2966},
   url={http://dx.doi.org/10.1093/mnras/stu1738},
   DOI={10.1093/mnras/stu1738},
   number={1},
   journal={Monthly Notices of the Royal Astronomical Society},
   publisher={Oxford University Press (OUP)},
   author={Hopkins, Philip F. and Kereš, Dušan and Oñorbe, José and Faucher-Giguère, Claude-André and Quataert, Eliot and Murray, Norman and Bullock, James S.},
   year={2014},
   month=sep, pages={581–603} }

@article{Weisz_2011,
   title={MODELING THE EFFECTS OF STAR FORMATION HISTORIES ON Hα AND ULTRAVIOLET FLUXES IN NEARBY DWARF GALAXIES},
   volume={744},
   ISSN={1538-4357},
   url={http://dx.doi.org/10.1088/0004-637X/744/1/44},
   DOI={10.1088/0004-637x/744/1/44},
   number={1},
   journal={The Astrophysical Journal},
   publisher={American Astronomical Society},
   author={Weisz, Daniel R. and Johnson, Benjamin D. and Johnson, L. Clifton and Skillman, Evan D. and Lee, Janice C. and Kennicutt, Robert C. and Calzetti, Daniela and van Zee, Liese and Bothwell, Matthew S. and Dalcanton, Julianne J. and Dale, Daniel A. and Williams, Benjamin F.},
   year={2011},
   month=dec, pages={44} }

@article{Tacchella_2016,
   title={Evolution of density profiles in high-zgalaxies: compaction and quenching inside-out},
   volume={458},
   ISSN={1365-2966},
   url={http://dx.doi.org/10.1093/mnras/stw303},
   DOI={10.1093/mnras/stw303},
   number={1},
   journal={Monthly Notices of the Royal Astronomical Society},
   publisher={Oxford University Press (OUP)},
   author={Tacchella, Sandro and Dekel, Avishai and Carollo, C. Marcella and Ceverino, Daniel and DeGraf, Colin and Lapiner, Sharon and Mandelker, Nir and Primack, Joel R.},
   year={2016},
   month=feb, pages={242–263} }

@article{Zolotov_2015,
   title={Compaction and quenching of high-z galaxies in cosmological simulations: blue and red nuggets},
   volume={450},
   ISSN={1365-2966},
   url={http://dx.doi.org/10.1093/mnras/stv740},
   DOI={10.1093/mnras/stv740},
   number={3},
   journal={Monthly Notices of the Royal Astronomical Society},
   publisher={Oxford University Press (OUP)},
   author={Zolotov, Adi and Dekel, Avishai and Mandelker, Nir and Tweed, Dylan and Inoue, Shigeki and DeGraf, Colin and Ceverino, Daniel and Primack, Joel R. and Barro, Guillermo and Faber, Sandra M.},
   year={2015},
   month=may, pages={2327–2353} }

@article{Tacchella_2015,
   title={Evidence for mature bulges and an inside-out quenching phase 3 billion years after the Big Bang},
   volume={348},
   ISSN={1095-9203},
   url={http://dx.doi.org/10.1126/science.1261094},
   DOI={10.1126/science.1261094},
   number={6232},
   journal={Science},
   publisher={American Association for the Advancement of Science (AAAS)},
   author={Tacchella, S. and Carollo, C. M. and Renzini, A. and Schreiber, N. M. Förster and Lang, P. and Wuyts, S. and Cresci, G. and Dekel, A. and Genzel, R. and Lilly, S. J. and Mancini, C. and Newman, S. and Onodera, M. and Shapley, A. and Tacconi, L. and Woo, J. and Zamorani, G.},
   year={2015},
   month=apr, pages={314–317} }

@article{Nelson_2016,
   title={WHERE STARS FORM: INSIDE-OUT GROWTH AND COHERENT STAR FORMATION FROM HST Hα MAPS OF 3200 GALAXIES ACROSS THE MAIN SEQUENCE AT 0.7 &lt; z &lt; 1.5},
   volume={828},
   ISSN={1538-4357},
   url={http://dx.doi.org/10.3847/0004-637X/828/1/27},
   DOI={10.3847/0004-637x/828/1/27},
   number={1},
   journal={The Astrophysical Journal},
   publisher={American Astronomical Society},
   author={Nelson, Erica June and van Dokkum, Pieter G. and Förster Schreiber, Natascha M. and Franx, Marijn and Brammer, Gabriel B. and Momcheva, Ivelina G. and Wuyts, Stijn and Whitaker, Katherine E. and Skelton, Rosalind E. and Fumagalli, Mattia and Hayward, Christopher C. and Kriek, Mariska and Labbé, Ivo and Leja, Joel and Rix, Hans-Walter and Tacconi, Linda J. and Wel, Arjen van der and Bosch, Frank C. van den and Oesch, Pascal A. and Dickey, Claire and Lange, Johannes Ulf},
   year={2016},
   month=aug, pages={27} }

@article{Bruzual_2003,
   title={Stellar population synthesis at the resolution of 2003},
   volume={344},
   ISSN={1365-2966},
   url={http://dx.doi.org/10.1046/j.1365-8711.2003.06897.x},
   DOI={10.1046/j.1365-8711.2003.06897.x},
   number={4},
   journal={Monthly Notices of the Royal Astronomical Society},
   publisher={Oxford University Press (OUP)},
   author={Bruzual, G. and Charlot, S.},
   year={2003},
   month=oct, pages={1000–1028} }

@article{Faucher_Gigu_re_2017,
   title={A model for the origin of bursty star formation in galaxies},
   volume={473},
   ISSN={1365-2966},
   url={http://dx.doi.org/10.1093/mnras/stx2595},
   DOI={10.1093/mnras/stx2595},
   number={3},
   journal={Monthly Notices of the Royal Astronomical Society},
   publisher={Oxford University Press (OUP)},
   author={Faucher-Giguère, Claude-André},
   year={2017},
   month=oct, pages={3717–3731} }

@article{Kriek_2015,
   title={THE MOSFIRE DEEP EVOLUTION FIELD (MOSDEF) SURVEY: REST-FRAME OPTICAL SPECTROSCOPY FOR ∼1500
                    H
                    -SELECTED GALAXIES AT $1.37\leqslant z\leqslant 3.8$},
   volume={218},
   ISSN={1538-4365},
   url={http://dx.doi.org/10.1088/0067-0049/218/2/15},
   DOI={10.1088/0067-0049/218/2/15},
   number={2},
   journal={The Astrophysical Journal Supplement Series},
   publisher={American Astronomical Society},
   author={Kriek, Mariska and Shapley, Alice E. and Reddy, Naveen A. and Siana, Brian and Coil, Alison L. and Mobasher, Bahram and Freeman, William R. and Groot, Laura de and Price, Sedona H. and Sanders, Ryan and Shivaei, Irene and Brammer, Gabriel B. and Momcheva, Ivelina G. and Skelton, Rosalind E. and Dokkum, Pieter G. van and Whitaker, Katherine E. and Aird, James and Azadi, Mojegan and Kassis, Marc and Bullock, James S. and Conroy, Charlie and Davé, Romeel and Kereš, Dušan and Krumholz, Mark},
   year={2015},
   month=may, pages={15} }

@article{Scholtz_2025,
   title={JADES: A large population of obscured, narrow-line active galactic nuclei at high redshift},
   volume={697},
   ISSN={1432-0746},
   url={http://dx.doi.org/10.1051/0004-6361/202348804},
   DOI={10.1051/0004-6361/202348804},
   journal={Astronomy \& Astrophysics},
   publisher={EDP Sciences},
   author={Scholtz, Jan and Maiolino, Roberto and D’Eugenio, Francesco and Curtis-Lake, Emma and Carniani, Stefano and Charlot, Stephane and Curti, Mirko and Silcock, Maddie S. and Arribas, Santiago and Baker, William and Bhatawdekar, Rachana and Boyett, Kristan and Bunker, Andrew J. and Chevallard, Jacopo and Circosta, Chiara and Eisenstein, Daniel J. and Hainline, Kevin and Hausen, Ryan and Ji, Xihan and Ji, Zhiyuan and Johnson, Benjamin D. and Kumari, Nimisha and Looser, Tobias J. and Lyu, Jianwei and Maseda, Michael V. and Parlanti, Eleonora and Perna, Michele and Rieke, Marcia and Robertson, Brant and Del Pino, Bruno Rodríguez and Sun, Fengwu and Tacchella, Sandro and Übler, Hannah and Venturi, Giacomo and Williams, Christina C. and Willmer, Christopher N. A. and Willott, Chris and Witstok, Joris},
   year={2025},
   month=may, pages={A175} }

@article{Vargas_2014,
   title={TO STACK OR NOT TO STACK: SPECTRAL ENERGY DISTRIBUTION PROPERTIES OF Lyα-EMITTING GALAXIES ATz= 2.1},
   volume={783},
   ISSN={1538-4357},
   url={http://dx.doi.org/10.1088/0004-637X/783/1/26},
   DOI={10.1088/0004-637x/783/1/26},
   number={1},
   journal={The Astrophysical Journal},
   publisher={American Astronomical Society},
   author={Vargas, Carlos J. and Bish, Hannah and Acquaviva, Viviana and Gawiser, Eric and Finkelstein, Steven L. and Ciardullo, Robin and Ashby, Matthew L. N. and Feldmeier, John and Ferguson, Henry and Gronwall, Caryl and Guaita, Lucia and Hagen, Alex and Koekemoer, Anton and Kurczynski, Peter and Newman, Jeffrey A. and Padilla, Nelson},
   year={2014},
   month=feb, pages={26} }

@article{Kusakabe_2015,
   title={FIRST INFRARED-BASED IMPLICATIONS FOR THE DUST ATTENUATION AND STAR FORMATION OF TYPICAL Ly
                    α
                    EMITTERS},
   volume={800},
   ISSN={2041-8213},
   url={http://dx.doi.org/10.1088/2041-8205/800/2/L29},
   DOI={10.1088/2041-8205/800/2/l29},
   number={2},
   journal={The Astrophysical Journal},
   publisher={American Astronomical Society},
   author={Kusakabe, Haruka and Shimasaku, Kazuhiro and Nakajima, Kimihiko and Ouchi, Masami},
   year={2015},
   month=feb, pages={L29} }

@ARTICLE{1986ApJ...307..431D,
       author = {{Dopita}, M.~A. and {Evans}, I.~N.},
        title = "{Theoretical Models for H II Regions. II. The Extragalactic H II Region Abundance Sequence}",
      journal = {\apj},
         year = 1986,
        month = aug,
       volume = {307},
        pages = {431},
          doi = {10.1086/164432},
       adsurl = {https://ui.adsabs.harvard.edu/abs/1986ApJ...307..431D}
}

@ARTICLE{2014MNRAS.441.2663P,
       author = {{P{\'e}rez-Montero}, E.},
        title = "{Deriving model-based T$_{e}$-consistent chemical abundances in ionized gaseous nebulae}",
      journal = {\mnras},
         year = 2014,
        month = jul,
       volume = {441},
       number = {3},
        pages = {2663-2675},
          doi = {10.1093/mnras/stu753},
archivePrefix = {arXiv},
       eprint = {1404.3936},
 primaryClass = {astro-ph.GA},
       adsurl = {https://ui.adsabs.harvard.edu/abs/2014MNRAS.441.2663P}
}

@ARTICLE{2009ApJ...700..221K,
       author = {{Kriek}, Mariska and {van Dokkum}, Pieter G. and {Labb{\'e}}, Ivo and {Franx}, Marijn and {Illingworth}, Garth D. and {Marchesini}, Danilo and {Quadri}, Ryan F.},
        title = "{An Ultra-Deep Near-Infrared Spectrum of a Compact Quiescent Galaxy at z = 2.2}",
      journal = {\apj},
         year = 2009,
        month = jul,
       volume = {700},
       number = {1},
        pages = {221-231},
          doi = {10.1088/0004-637X/700/1/221},
archivePrefix = {arXiv},
       eprint = {0905.1692},
 primaryClass = {astro-ph.CO},
       adsurl = {https://ui.adsabs.harvard.edu/abs/2009ApJ...700..221K}
}

@ARTICLE{2015ApJ...801...88S,
       author = {{Shapley}, Alice E. and {Reddy}, Naveen A. and {Kriek}, Mariska and {Freeman}, William R. and {Sanders}, Ryan L. and {Siana}, Brian and {Coil}, Alison L. and {Mobasher}, Bahram and {Shivaei}, Irene and {Price}, Sedona H. and {de Groot}, Laura},
        title = "{The MOSDEF Survey: Excitation Properties of z {\ensuremath{\sim}} 2.3 Star-forming Galaxies}",
      journal = {\apj},
         year = 2015,
        month = mar,
       volume = {801},
       number = {2},
          eid = {88},
        pages = {88},
          doi = {10.1088/0004-637X/801/2/88},
archivePrefix = {arXiv},
       eprint = {1409.7071},
 primaryClass = {astro-ph.GA},
       adsurl = {https://ui.adsabs.harvard.edu/abs/2015ApJ...801...88S}
}

@misc{jin2025highincidencecentralstar,
      title={A High Incidence of Central Star Formation Inferred from the Color Gradients of Galaxies at $z>4$}, 
      author={Bingcheng Jin and Luis C. Ho and Wen Sun},
      year={2025},
      eprint={2412.03455},
      archivePrefix={arXiv},
      primaryClass={astro-ph.GA},
      url={https://arxiv.org/abs/2412.03455}, 
}

@ARTICLE{2012ApJ...745...12N,
       author = {{Nakajima}, Kimihiko and {Ouchi}, Masami and {Shimasaku}, Kazuhiro and {Ono}, Yoshiaki and {Lee}, Janice C. and {Foucaud}, Sebastien and {Ly}, Chun and {Dale}, Daniel A. and {Salim}, Samir and {Finn}, Rose and {Almaini}, Omar and {Okamura}, Sadanori},
        title = "{Average Metallicity and Star Formation Rate of Ly{\ensuremath{\alpha}} Emitters Probed by a Triple Narrowband Survey}",
      journal = {\apj},
         year = 2012,
        month = jan,
       volume = {745},
       number = {1},
          eid = {12},
        pages = {12},
          doi = {10.1088/0004-637X/745/1/12},
archivePrefix = {arXiv},
       eprint = {1105.2824},
 primaryClass = {astro-ph.CO},
       adsurl = {https://ui.adsabs.harvard.edu/abs/2012ApJ...745...12N}
}

@ARTICLE{2018PASJ...70....4K,
       author = {{Kusakabe}, Haruka and {Shimasaku}, Kazuhiro and {Ouchi}, Masami and {Nakajima}, Kimihiko and {Goto}, Ryosuke and {Hashimoto}, Takuya and {Konno}, Akira and {Harikane}, Yuichi and {Silverman}, John D. and {Capak}, Peter L.},
        title = "{The stellar mass, star formation rate and dark matter halo properties of LAEs at z {\ensuremath{\sim}} 2}",
      journal = {\pasj},
         year = 2018,
        month = jan,
       volume = {70},
       number = {1},
          eid = {4},
        pages = {4},
          doi = {10.1093/pasj/psx148},
archivePrefix = {arXiv},
       eprint = {1707.09373},
 primaryClass = {astro-ph.GA},
       adsurl = {https://ui.adsabs.harvard.edu/abs/2018PASJ...70....4K}
}

@ARTICLE{2011ApJ...733..114G,
       author = {{Guaita}, Lucia and {Acquaviva}, Viviana and {Padilla}, Nelson and {Gawiser}, Eric and {Bond}, Nicholas A. and {Ciardullo}, Robin and {Treister}, Ezequiel and {Kurczynski}, Peter and {Gronwall}, Caryl and {Lira}, Paulina and {Schawinski}, Kevin},
        title = "{Ly{\ensuremath{\alpha}}-emitting Galaxies at z = 2.1: Stellar Masses, Dust, and Star Formation Histories from Spectral Energy Distribution Fitting}",
      journal = {\apj},
         year = 2011,
        month = jun,
       volume = {733},
       number = {2},
          eid = {114},
        pages = {114},
          doi = {10.1088/0004-637X/733/2/114},
archivePrefix = {arXiv},
       eprint = {1101.3017},
 primaryClass = {astro-ph.CO},
       adsurl = {https://ui.adsabs.harvard.edu/abs/2011ApJ...733..114G}
}

@ARTICLE{2011A&A...529A...9N,
       author = {{Nilsson}, K.~K. and {{\"O}stlin}, G. and {M{\o}ller}, P. and {M{\"o}ller-Nilsson}, O. and {Tapken}, C. and {Freudling}, W. and {Fynbo}, J.~P.~U.},
        title = "{The nature of z \raisebox{-0.5ex}\textasciitilde 2.3 Lyman-{\ensuremath{\alpha}} emitters}",
      journal = {\aap},
         year = 2011,
        month = may,
       volume = {529},
          eid = {A9},
        pages = {A9},
          doi = {10.1051/0004-6361/201015606},
archivePrefix = {arXiv},
       eprint = {1009.0007},
 primaryClass = {astro-ph.CO},
       adsurl = {https://ui.adsabs.harvard.edu/abs/2011A&A...529A...9N}
}

@ARTICLE{2023A&A...677A.115C,
       author = {{Cameron}, Alex J. and {Saxena}, Aayush and {Bunker}, Andrew J. and {D'Eugenio}, Francesco and {Carniani}, Stefano and {Maiolino}, Roberto and {Curtis-Lake}, Emma and {Ferruit}, Pierre and {Jakobsen}, Peter and {Arribas}, Santiago and {Bonaventura}, Nina and {Charlot}, Stephane and {Chevallard}, Jacopo and {Curti}, Mirko and {Looser}, Tobias J. and {Maseda}, Michael V. and {Rawle}, Tim and {Rodr{\'\i}guez Del Pino}, Bruno and {Smit}, Renske and {{\"U}bler}, Hannah and {Willott}, Chris and {Witstok}, Joris and {Egami}, Eiichi and {Eisenstein}, Daniel J. and {Johnson}, Benjamin D. and {Hainline}, Kevin and {Rieke}, Marcia and {Robertson}, Brant E. and {Stark}, Daniel P. and {Tacchella}, Sandro and {Williams}, Christina C. and {Willmer}, Christopher N.~A. and {Bhatawdekar}, Rachana and {Bowler}, Rebecca and {Boyett}, Kristan and {Circosta}, Chiara and {Helton}, Jakob M. and {Jones}, Gareth C. and {Kumari}, Nimisha and {Ji}, Zhiyuan and {Nelson}, Erica and {Parlanti}, Eleonora and {Sandles}, Lester and {Scholtz}, Jan and {Sun}, Fengwu},
        title = "{JADES: Probing interstellar medium conditions at z {\ensuremath{\sim}} 5.5-9.5 with ultra-deep JWST/NIRSpec spectroscopy}",
      journal = {\aap},
         year = 2023,
        month = sep,
       volume = {677},
          eid = {A115},
        pages = {A115},
          doi = {10.1051/0004-6361/202346107},
archivePrefix = {arXiv},
       eprint = {2302.04298},
 primaryClass = {astro-ph.GA},
       adsurl = {https://ui.adsabs.harvard.edu/abs/2023A&A...677A.115C}
}

@ARTICLE{2017ApJ...847...38Y,
       author = {{Yang}, Huan and {Malhotra}, Sangeeta and {Rhoads}, James E. and {Wang}, Junxian},
        title = "{Blueberry Galaxies: The Lowest Mass Young Starbursts}",
      journal = {\apj},
         year = 2017,
        month = sep,
       volume = {847},
       number = {1},
          eid = {38},
        pages = {38},
          doi = {10.3847/1538-4357/aa8809},
archivePrefix = {arXiv},
       eprint = {1706.02819},
 primaryClass = {astro-ph.GA},
       adsurl = {https://ui.adsabs.harvard.edu/abs/2017ApJ...847...38Y}
}

@ARTICLE{2017ApJ...844..171Y,
       author = {{Yang}, Huan and {Malhotra}, Sangeeta and {Gronke}, Max and {Rhoads}, James E. and {Leitherer}, Claus and {Wofford}, Aida and {Jiang}, Tianxing and {Dijkstra}, Mark and {Tilvi}, V. and {Wang}, Junxian},
        title = "{Ly{\ensuremath{\alpha}} Profile, Dust, and Prediction of Ly{\ensuremath{\alpha}} Escape Fraction in Green Pea Galaxies}",
      journal = {\apj},
         year = 2017,
        month = aug,
       volume = {844},
       number = {2},
          eid = {171},
        pages = {171},
          doi = {10.3847/1538-4357/aa7d4d},
archivePrefix = {arXiv},
       eprint = {1701.01857},
 primaryClass = {astro-ph.GA},
       adsurl = {https://ui.adsabs.harvard.edu/abs/2017ApJ...844..171Y}
}

@article{Terao_2022,
   title={A Selection of Hα Emitters at z = 2.1–2.5 Using the K
               
                  s
               -band Photometry of ZFOURGE},
   volume={941},
   ISSN={1538-4357},
   url={http://dx.doi.org/10.3847/1538-4357/ac9fce},
   DOI={10.3847/1538-4357/ac9fce},
   number={1},
   journal={The Astrophysical Journal},
   publisher={American Astronomical Society},
   author={Terao, Yasunori and Spitler, Lee R. and Motohara, Kentaro and Chen, Nuo},
   year={2022},
   month=dec, pages={70} }

@article{Dutton_2010,
   title={On the origin of the galaxy star-formation-rate sequence: evolution and scatter: Origin of the galaxy SFR sequence},
   ISSN={1365-2966},
   url={http://dx.doi.org/10.1111/j.1365-2966.2010.16620.x},
   DOI={10.1111/j.1365-2966.2010.16620.x},
   journal={Monthly Notices of the Royal Astronomical Society},
   publisher={Oxford University Press (OUP)},
   author={Dutton, Aaron A. and Van Den Bosch, Frank C. and Dekel, Avishai},
   year={2010},
   month=apr, pages={no-no} }

@article{Matthee_2019,
   title={The origin of scatter in the star formation rate–stellar mass relation},
   volume={484},
   ISSN={1365-2966},
   url={http://dx.doi.org/10.1093/mnras/stz030},
   DOI={10.1093/mnras/stz030},
   number={1},
   journal={Monthly Notices of the Royal Astronomical Society},
   publisher={Oxford University Press (OUP)},
   author={Matthee, Jorryt and Schaye, Joop},
   year={2019},
   month=jan, pages={915–932} }

@ARTICLE{2025MNRAS.541.1329D,
       author = {{Duncan}, K.~J. and {McLeod}, D.~J. and {Best}, P.~N. and {Pirie}, C.~A. and {Clausen}, M. and {Cochrane}, R.~K. and {Dunlop}, J.~S. and {Flury}, S.~R. and {Geach}, J.~E. and {Grogin}, N.~A. and {Hale}, C.~L. and {Ibar}, E. and {Kondapally}, R. and {Li}, Zefeng and {Matthee}, J. and {McLure}, R.~J. and {Ossa-Fuentes}, Luis and {Patrick}, A.~L. and {Smail}, Ian and {Sobral}, D. and {Stephenson}, H.~M.~O. and {Stott}, J.~P. and {Swinbank}, A.~M.},
        title = "{The JWST Emission-Line Survey: extending rest-optical narrow-band emission-line selection into the Epoch of Reionization}",
      journal = {\mnras},
         year = 2025,
        month = aug,
       volume = {541},
       number = {2},
        pages = {1329-1347},
          doi = {10.1093/mnras/staf1061},
archivePrefix = {arXiv},
       eprint = {2410.09000},
 primaryClass = {astro-ph.GA},
       adsurl = {https://ui.adsabs.harvard.edu/abs/2025MNRAS.541.1329D}
}

@ARTICLE{2025MNRAS.541.1348P,
       author = {{Pirie}, C.~A. and {Best}, P.~N. and {Duncan}, K.~J. and {McLeod}, D.~J. and {Cochrane}, R.~K. and {Clausen}, M. and {Dunlop}, J.~S. and {Flury}, S.~R. and {Geach}, J.~E. and {Hale}, C.~L. and {Ibar}, E. and {Kondapally}, R. and {Li}, Zefeng and {Matthee}, J. and {McLure}, R.~J. and {Ossa-Fuentes}, L. and {Patrick}, A.~L. and {Smail}, Ian and {Sobral}, D. and {Stephenson}, H.~M.~O. and {Stott}, J.~P. and {Swinbank}, A.~M.},
        title = "{The JWST Emission Line Survey (JELS): an untargeted search for H {\ensuremath{\alpha}} emission line galaxies at z > 6 and their physical properties}",
      journal = {\mnras},
         year = 2025,
        month = aug,
       volume = {541},
       number = {2},
        pages = {1348-1376},
          doi = {10.1093/mnras/staf1006},
archivePrefix = {arXiv},
       eprint = {2410.11808},
 primaryClass = {astro-ph.GA},
       adsurl = {https://ui.adsabs.harvard.edu/abs/2025MNRAS.541.1348P}
}

@MISC{2021jwst.prop.1837D,
       author = {{Dunlop}, James S. and {Abraham}, Roberto G. and {Ashby}, Matthew L.~N. and {Bagley}, Micaela and {Best}, Philip N. and {Bongiorno}, Angela and {Bouwens}, Rychard and {Bowler}, Rebecca A.~A. and {Brammer}, Gabriel and {Bremer}, Malcolm and {Calabro'}, Antonello and {Carnall}, Adam and {Castellano}, Marco and {Cirasuolo}, Michele and {Conselice}, Christopher and {Cullen}, Fergus and {Dave}, Romeel and {Dayal}, Pratika and {Dekel}, Avishai and {Dickinson}, Mark and {Duncan}, Kenneth James and {Elbaz}, David and {Ellis}, Richard S. and {Ferguson}, Harry C. and {Ferrara}, Andrea and {Finkelstein}, Steven L. and {Fontana}, Adriano and {Furlanetto}, Steven and {Fynbo}, Johan P.~U. and {Gallerani}, Simona and {Gardner}, Jonathan P. and {Giavalisco}, Mauro and {Grazian}, Andrea and {Grogin}, Norman and {Harikane}, Yuichi and {Hopkins}, Philip F. and {Ilbert}, Olivier and {Illingworth}, Garth D. and {Juneau}, Stephanie and {Jung}, Intae and {Kartaltepe}, Jeyhan and {Kassin}, Susan and {Kauffmann}, Olivier Benjamin and {Khochfar}, Sadegh and {Kirkpatrick}, Allison and {Kocevski}, Dale D. and {Koekemoer}, Anton M. and {Labbe}, Ivo and {Laporte}, Nicolas and {Larson}, Rebecca L. and {Lucas}, Ray A. and {Magee}, Daniel K. and {Mason}, Charlotte and {McCracken}, Henry Joy and {McLeod}, Derek and {McLure}, Ross and {Merlin}, Emiliano and {Mesinger}, Andrei and {Milvang-Jensen}, Bo and {Newman}, Jeffrey Allen and {Oesch}, Pascal and {Ouchi}, Masami and {Pacifici}, Camilla and {Papovich}, Casey and {Peacock}, John and {Peeples}, Molly and {Pentericci}, Laura and {Perez-Gonzalez}, Pablo G. and {Pirzkal}, Norbert and {Pope}, Alexandra and {Pye}, John P. and {Reddy}, Naveen A. and {Robertson}, Brant and {Salvato}, Mara and {Santini}, Paola and {Schaerer}, Daniel and {Shapley}, Alice E. and {Simons}, Raymond and {Smit}, Renske and {Smith}, Britton D. and {Snyder}, Greg and {Somerville}, Rachel S. and {Stanway}, Elizabeth R. and {Stefanon}, Mauro and {Tasca}, Lidia and {Tikkanen}, Tuomo and {Tresse}, Laurence and {Trump}, Jonathan R. and {Whitaker}, Katherine E. and {Wilkins}, Stephen Matthew and {Wright}, Gillian and {Wyithe}, J. Stuart B. and {van Dokkum}, Pieter and {van der Werf}, Paul},
        title = "{PRIMER: Public Release IMaging for Extragalactic Research}",
 howpublished = {JWST Proposal. Cycle 1, ID. \#1837},
         year = 2021,
        month = mar,
        pages = {1837},
       adsurl = {https://ui.adsabs.harvard.edu/abs/2021jwst.prop.1837D}
}

@ARTICLE{2004MNRAS.348L..59P,
       author = {{Pettini}, Max and {Pagel}, Bernard E.~J.},
        title = "{[OIII]/[NII] as an abundance indicator at high redshift}",
      journal = {\mnras},
         year = 2004,
        month = mar,
       volume = {348},
       number = {3},
        pages = {L59-L63},
          doi = {10.1111/j.1365-2966.2004.07591.x},
archivePrefix = {arXiv},
       eprint = {astro-ph/0401128},
 primaryClass = {astro-ph},
       adsurl = {https://ui.adsabs.harvard.edu/abs/2004MNRAS.348L..59P}
}

@ARTICLE{2013A&A...559A.114M,
       author = {{Marino}, R.~A. and {Rosales-Ortega}, F.~F. and {S{\'a}nchez}, S.~F. and {Gil de Paz}, A. and {V{\'\i}lchez}, J. and {Miralles-Caballero}, D. and {Kehrig}, C. and {P{\'e}rez-Montero}, E. and {Stanishev}, V. and {Iglesias-P{\'a}ramo}, J. and {D{\'\i}az}, A.~I. and {Castillo-Morales}, A. and {Kennicutt}, R. and {L{\'o}pez-S{\'a}nchez}, A.~R. and {Galbany}, L. and {Garc{\'\i}a-Benito}, R. and {Mast}, D. and {Mendez-Abreu}, J. and {Monreal-Ibero}, A. and {Husemann}, B. and {Walcher}, C.~J. and {Garc{\'\i}a-Lorenzo}, B. and {Masegosa}, J. and {Del Olmo Orozco}, A. and {Mour{\~a}o}, A.~M. and {Ziegler}, B. and {Moll{\'a}}, M. and {Papaderos}, P. and {S{\'a}nchez-Bl{\'a}zquez}, P. and {Gonz{\'a}lez Delgado}, R.~M. and {Falc{\'o}n-Barroso}, J. and {Roth}, M.~M. and {van de Ven}, G. and {CALIFA Team}},
        title = "{The O3N2 and N2 abundance indicators revisited: improved calibrations based on CALIFA and T$_{e}$-based literature data}",
      journal = {\aap},
         year = 2013,
        month = nov,
       volume = {559},
          eid = {A114},
        pages = {A114},
          doi = {10.1051/0004-6361/201321956},
archivePrefix = {arXiv},
       eprint = {1307.5316},
 primaryClass = {astro-ph.CO},
       adsurl = {https://ui.adsabs.harvard.edu/abs/2013A&A...559A.114M}
}

@ARTICLE{2015PASA...32...27H,
       author = {{Hayes}, Matthew},
        title = "{Lyman Alpha Emitting Galaxies in the Nearby Universe}",
      journal = {\pasa},
         year = 2015,
        month = jul,
       volume = {32},
          eid = {e027},
        pages = {e027},
          doi = {10.1017/pasa.2015.25},
archivePrefix = {arXiv},
       eprint = {1505.07483},
 primaryClass = {astro-ph.GA},
       adsurl = {https://ui.adsabs.harvard.edu/abs/2015PASA...32...27H}
}

@ARTICLE{2016A&A...586A..83P,
       author = {{Puglisi}, A. and {Rodighiero}, G. and {Franceschini}, A. and {Talia}, M. and {Cimatti}, A. and {Baronchelli}, I. and {Daddi}, E. and {Renzini}, A. and {Schawinski}, K. and {Mancini}, C. and et al.},
        title = "{Dust attenuation in z \raisebox{-0.5ex}\textasciitilde 1 galaxies from Herschel and 3D-HST H{\ensuremath{\alpha}} measurements}",
      journal = {\aap},
         year = 2016,
        month = feb,
       volume = {586},
          eid = {A83},
        pages = {A83},
          doi = {10.1051/0004-6361/201526782},
archivePrefix = {arXiv},
       eprint = {1507.00005},
 primaryClass = {astro-ph.GA},
       adsurl = {https://ui.adsabs.harvard.edu/abs/2016A&A...586A..83P}
}

@ARTICLE{1981PASP...93....5B,
       author = {{Baldwin}, J.~A. and {Phillips}, M.~M. and {Terlevich}, R.},
        title = "{Classification parameters for the emission-line spectra of extragalactic objects.}",
      journal = {\pasp},
         year = 1981,
        month = feb,
       volume = {93},
        pages = {5-19},
          doi = {10.1086/130766},
       adsurl = {https://ui.adsabs.harvard.edu/abs/1981PASP...93....5B}
}

@misc{sanders2023excitationionizationpropertiesstarforming,
      title={Excitation and Ionization Properties of Star-forming Galaxies at z=2.0-9.3 with JWST/NIRSpec}, 
      author={Ryan L. Sanders and Alice E. Shapley and Michael W. Topping and Naveen A. Reddy and Gabriel B. Brammer},
      year={2023},
      eprint={2301.06696},
      archivePrefix={arXiv},
      primaryClass={astro-ph.GA},
      url={https://arxiv.org/abs/2301.06696}, 
}

@ARTICLE{2023A&A...677A.114A,
       author = {{Arroyo-Polonio}, A. and {Iglesias-P{\'a}ramo}, J. and {Kehrig}, C. and {V{\'\i}lchez}, J.~M. and {Amor{\'\i}n}, R. and {Breda}, I. and {P{\'e}rez-Montero}, E. and {P{\'e}rez-D{\'\i}az}, B. and {Hayes}, M.},
        title = "{A MUSE/VLT spatially resolved study of the emission structure of Green Pea galaxies}",
      journal = {\aap},
         year = 2023,
        month = sep,
       volume = {677},
          eid = {A114},
        pages = {A114},
          doi = {10.1051/0004-6361/202346192},
archivePrefix = {arXiv},
       eprint = {2309.09585},
 primaryClass = {astro-ph.GA},
       adsurl = {https://ui.adsabs.harvard.edu/abs/2023A&A...677A.114A}
}

@ARTICLE{2024A&A...688A.159K,
       author = {{Kouroumpatzakis}, K. and {Svoboda}, J. and {Zezas}, A. and {Borkar}, A. and {Kyritsis}, E. and {Boorman}, P.~G. and {Daoutis}, C. and {Adamcov{\'a}}, B. and {Grossov{\'a}}, R.},
        title = "{Blueberry galaxies up to 200 Mpc and their optical and infrared properties}",
      journal = {\aap},
         year = 2024,
        month = aug,
       volume = {688},
          eid = {A159},
        pages = {A159},
          doi = {10.1051/0004-6361/202449766},
archivePrefix = {arXiv},
       eprint = {2405.03391},
 primaryClass = {astro-ph.GA},
       adsurl = {https://ui.adsabs.harvard.edu/abs/2024A&A...688A.159K}
}

@ARTICLE{2016ApJ...828...27N,
       author = {{Nelson}, Erica June and {van Dokkum}, Pieter G. and {F{\"o}rster Schreiber}, Natascha M. and {Franx}, Marijn and {Brammer}, Gabriel B. and {Momcheva}, Ivelina G. and {Wuyts}, Stijn and {Whitaker}, Katherine E. and {Skelton}, Rosalind E. and {Fumagalli}, Mattia and {Hayward}, Christopher C. and {Kriek}, Mariska and {Labb{\'e}}, Ivo and {Leja}, Joel and {Rix}, Hans-Walter and {Tacconi}, Linda J. and {van der Wel}, Arjen and {van den Bosch}, Frank C. and {Oesch}, Pascal A. and {Dickey}, Claire and {Ulf Lange}, Johannes},
        title = "{Where Stars Form: Inside-out Growth and Coherent Star Formation from HST H{\ensuremath{\alpha}} Maps of 3200 Galaxies across the Main Sequence at 0.7 < z < 1.5}",
      journal = {\apj},
         year = 2016,
        month = sep,
       volume = {828},
       number = {1},
          eid = {27},
        pages = {27},
          doi = {10.3847/0004-637X/828/1/27},
archivePrefix = {arXiv},
       eprint = {1507.03999},
 primaryClass = {astro-ph.GA},
       adsurl = {https://ui.adsabs.harvard.edu/abs/2016ApJ...828...27N}
}

@ARTICLE{2019PASJ...71...69S,
       author = {{Suzuki}, Tomoko L. and {Minowa}, Yosuke and {Koyama}, Yusei and {Kodama}, Tadayuki and {Hayashi}, Masao and {Shimakawa}, Rhythm and {Tanaka}, Ichi and {Tadaki}, Ken-ichi},
        title = "{Extended star-forming regions within galaxies in a dense proto-cluster core at z = 2.53}",
      journal = {\pasj},
         year = 2019,
        month = aug,
       volume = {71},
       number = {4},
          eid = {69},
        pages = {69},
          doi = {10.1093/pasj/psz047},
archivePrefix = {arXiv},
       eprint = {1904.00695},
 primaryClass = {astro-ph.GA},
       adsurl = {https://ui.adsabs.harvard.edu/abs/2019PASJ...71...69S}
}

@ARTICLE{2012ApJ...754..120M,
       author = {{Messias}, H. and {Afonso}, J. and {Salvato}, M. and {Mobasher}, B. and {Hopkins}, A.~M.},
        title = "{A New Infrared Color Criterion for the Selection of 0 < z < 7 AGNs: Application to Deep Fields and Implications for JWST Surveys}",
      journal = {\apj},
         year = 2012,
        month = aug,
       volume = {754},
       number = {2},
          eid = {120},
        pages = {120},
          doi = {10.1088/0004-637X/754/2/120},
archivePrefix = {arXiv},
       eprint = {1205.4764},
 primaryClass = {astro-ph.CO},
       adsurl = {https://ui.adsabs.harvard.edu/abs/2012ApJ...754..120M}
}

@ARTICLE{2007ApJ...663...81P,
       author = {{Polletta}, M. and {Tajer}, M. and {Maraschi}, L. and {Trinchieri}, G. and {Lonsdale}, C.~J. and {Chiappetti}, L. and {Andreon}, S. and {Pierre}, M. and {Le F{\`e}vre}, O. and {Zamorani}, G. and {Maccagni}, D. and {Garcet}, O. and {Surdej}, J. and {Franceschini}, A. and {Alloin}, D. and {Shupe}, D.~L. and {Surace}, J.~A. and {Fang}, F. and {Rowan-Robinson}, M. and {Smith}, H.~E. and {Tresse}, L.},
        title = "{Spectral Energy Distributions of Hard X-Ray Selected Active Galactic Nuclei in the XMM-Newton Medium Deep Survey}",
      journal = {\apj},
         year = 2007,
        month = jul,
       volume = {663},
       number = {1},
        pages = {81-102},
          doi = {10.1086/518113},
archivePrefix = {arXiv},
       eprint = {astro-ph/0703255},
 primaryClass = {astro-ph},
       adsurl = {https://ui.adsabs.harvard.edu/abs/2007ApJ...663...81P}
}

@ARTICLE{2014ApJ...790..113Z,
       author = {{Zeimann}, Gregory R. and {Ciardullo}, Robin and {Gebhardt}, Henry and {Gronwall}, Caryl and {Schneider}, Donald P. and {Hagen}, Alex and {Bridge}, Joanna S. and {Feldmeier}, John and {Trump}, Jonathan R.},
        title = "{3D-HST Emission Line Galaxies at z \raisebox{-0.5ex}\textasciitilde 2: Discrepancies in the Optical/UV Star Formation Rates}",
      journal = {\apj},
         year = 2014,
        month = aug,
       volume = {790},
       number = {2},
          eid = {113},
        pages = {113},
          doi = {10.1088/0004-637X/790/2/113},
archivePrefix = {arXiv},
       eprint = {1406.3355},
 primaryClass = {astro-ph.GA},
       adsurl = {https://ui.adsabs.harvard.edu/abs/2014ApJ...790..113Z}
}

@ARTICLE{2022MNRAS.509.4940T,
       author = {{Thorne}, Jessica E. and {Robotham}, Aaron S.~G. and {Davies}, Luke J.~M. and {Bellstedt}, Sabine and {Brown}, Michael J.~I. and {Croom}, Scott M. and {Delvecchio}, Ivan and {Groves}, Brent and {Jarvis}, Matt J. and {Shabala}, Stanislav S. and {Seymour}, Nick and {Whittam}, Imogen H. and {Bravo}, Matias and {Cook}, Robin H.~W. and {Driver}, Simon P. and {Holwerda}, Benne and {Phillipps}, Steven and {Siudek}, Malgorzata},
        title = "{Deep Extragalactic VIsible Legacy Survey (DEVILS): identification of AGN through SED fitting and the evolution of the bolometric AGN luminosity function}",
      journal = {\mnras},
         year = 2022,
        month = feb,
       volume = {509},
       number = {4},
        pages = {4940-4961},
          doi = {10.1093/mnras/stab3208},
archivePrefix = {arXiv},
       eprint = {2112.06366},
 primaryClass = {astro-ph.GA},
       adsurl = {https://ui.adsabs.harvard.edu/abs/2022MNRAS.509.4940T}
}

@INPROCEEDINGS{2020SPIE11447E..5HK,
       author = {{Konishi}, Masahiro and {Motohara}, Kentaro and {Takahashi}, Hidenori and {Kato}, Natsuko and {Kushibiki}, Kosuke and {Nakamura}, Hiroki and {Chen}, Nuo and {Kodama}, Tadayuki and {Hayashi}, Masao and {Tanaka}, Ichi and {Tadaki}, Ken-ichi and {Toshikawa}, Jun and {Koyama}, Yusei and {Shimakawa}, Rhythm and {Suzuki}, Tomoko L. and {Asano}, Tetsuro and {Okita}, Hirofumi and {Koshida}, Shintaro and {Tateuchi}, Ken and {Todo}, Soya and {Kobayakawa}, Yutaka and {Kitagawa}, Yutaro and {Ohashi}, Hirofumi and {Kono}, Yukihiro and {Terao}, Yasunori and {Aoki}, Tsutomu and {Asano}, Kentaro and {Doi}, Mamoru and {Hatsukade}, Bunyo and {Kamizuka}, Takafumi and {Kohno}, Kotaro and {Minezaki}, Takeo and {Miyata}, Takashi and {Morokuma}, Tomoki and {Numata}, Mizuki and {Sako}, Shigeyuki and {Sameshima}, Hiroaki and {Soyano}, Takao and {Tanab{\'e}}, Toshihiko and {Tanaka}, Masuo and {Tarusawa}, Ken'ichi and {Ohsawa}, Ryou and {Tachibana}, Kengo and {Michifuji}, Tsubasa and {Yoshii}, Yuzuru},
        title = "{The University of Tokyo Atacama Observatory 6.5m telescope: On-sky performance of the near-infrared instrument SWIMS on the Subaru telescope}",
    booktitle = {Ground-based and Airborne Instrumentation for Astronomy VIII},
         year = 2020,
       editor = {{Evans}, Christopher J. and {Bryant}, Julia J. and {Motohara}, Kentaro},
       series = {Society of Photo-Optical Instrumentation Engineers (SPIE) Conference Series},
       volume = {11447},
        month = dec,
          eid = {114475H},
        pages = {114475H},
          doi = {10.1117/12.2560422},
       adsurl = {https://ui.adsabs.harvard.edu/abs/2020SPIE11447E..5HK}
}

@ARTICLE{2000A&A...354.1134R,
       author = {{Rousselot}, P. and {Lidman}, C. and {Cuby}, J.-G. and {Moreels}, G. and {Monnet}, G.},
        title = "{Night-sky spectral atlas of OH emission lines in the near-infrared}",
      journal = {\aap},
         year = 2000,
        month = feb,
       volume = {354},
        pages = {1134-1150},
       adsurl = {https://ui.adsabs.harvard.edu/abs/2000A&A...354.1134R}
}

@ARTICLE{2024A&A...690A..64M,
       author = {{Matharu}, Jasleen and {Nelson}, Erica J. and {Brammer}, Gabriel and {Oesch}, Pascal A. and {Allen}, Natalie and {Shivaei}, Irene and {Naidu}, Rohan P. and {Chisholm}, John and {Covelo-Paz}, Alba and {Fudamoto}, Yoshinobu and {Giovinazzo}, Emma and {Herard-Demanche}, Thomas and {Kerutt}, Josephine and {Kramarenko}, Ivan and {Marchesini}, Danilo and {Meyer}, Romain A. and {Prieto-Lyon}, Gonzalo and {Reddy}, Naveen and {Shuntov}, Marko and {Weibel}, Andrea and {Wuyts}, Stijn and {Xiao}, Mengyuan},
        title = "{A first look at spatially resolved star formation at 4.8 < z < 6.5 with JWST FRESCO NIRCam slitless spectroscopy}",
      journal = {\aap},
         year = 2024,
        month = oct,
       volume = {690},
          eid = {A64},
        pages = {A64},
          doi = {10.1051/0004-6361/202450522},
archivePrefix = {arXiv},
       eprint = {2404.17629},
 primaryClass = {astro-ph.GA},
       adsurl = {https://ui.adsabs.harvard.edu/abs/2024A&A...690A..64M}
}

@ARTICLE{2019ApJ...876...12A,
       author = {{Agostino}, Christopher J. and {Salim}, Samir},
        title = "{Crossing the Line: Active Galactic Nuclei in the Star-forming Region of the BPT Diagram}",
      journal = {\apj},
         year = 2019,
        month = may,
       volume = {876},
       number = {1},
          eid = {12},
        pages = {12},
          doi = {10.3847/1538-4357/ab1094},
archivePrefix = {arXiv},
       eprint = {1811.11755},
 primaryClass = {astro-ph.GA},
       adsurl = {https://ui.adsabs.harvard.edu/abs/2019ApJ...876...12A}
}
\bibliographystyle{aasjournalv7}



\end{document}